\documentclass[a4paper,11pt]{article}
\pdfoutput=1 

\usepackage{jcappub} 

\usepackage[T1]{fontenc} 
\usepackage{graphicx}
\usepackage[export]{adjustbox}
\usepackage{amsmath}
\usepackage{array}
\usepackage{subcaption}
\usepackage{amsfonts}
\usepackage{amssymb}
\usepackage[left=2cm,right=2cm,top=2cm,bottom=2cm]{geometry}
\usepackage{array}
\usepackage{multirow}
\usepackage{multicol}
\usepackage{tablefootnote}

\def\aj{AJ}
\def\apj{ApJ}

\def\mnras{MNRAS}
\def\aap{A\&A}

\def\nat{Nature}      

\def\apjl{ApJ Letters}

\def\pasp{Publications of the Astronomical Society of the Pacific}

\title {\boldmath The correlations between galaxy properties in different environments of the cosmic web}
\author[a]{Anindita Nandi,}

\author[a]{Biswajit Pandey}

\author[b]{and Prakash Sarkar}

\affiliation[a]{Department of Physics, Visva-Bharati University, Santiniketan, 731235, India}
\affiliation[b]{Department of Physics, Kashi Sahu College, Seraikella, 833219, India}

\emailAdd{anindita.nandi96@gmail.com} \emailAdd{biswap@visva-bharati.ac.in} \emailAdd{prakash.sarkar@gmail.com} 

\abstract{We study the correlations between $(u-r)$
    colour, stellar mass, specific star formation rate (sSFR) and
    metallicity of galaxies in different geometric environments of the
    cosmic web using a volume limited sample from the SDSS. The
    geometric environment at the location of each galaxy is determined
    using the eigenvalues of the tidal tensor in three dimensions. We
    use the Pearson correlation coefficient (PCC) and the normalized
    mutual information (NMI) to quantify the correlations between
    these galaxy properties in sheets, filaments and clusters after
    matching the stellar mass distributions of the galaxies in these
    environments.  A two-tailed t-test assesses the statistical
    significance of the observed differences between these relations
    in different geometric environments. The null hypothesis can be
    rejected at $>99.99\%$ significance level in most of the cases,
    suggesting that the scaling relations between the observable
    galaxy properties are susceptible to the geometric environments of
    the cosmic web.}

\begin{document}
\maketitle
\flushbottom


\section{Introduction}
The galaxies are the fundamental units of the large-scale structures
in the Universe. A complete understanding of their formation and
evolution remains elusive. In the current paradigm, every galaxy is
believed to have formed within a dark matter halo. The dark matter
halos form from the gravitational collapse of density fluctuations and
continue to grow through hierarchical merging. The halos accrete gas
from the IGM that radiatively cool and settle down at their centres to
form galaxies \citep{reesostriker77, silk77, white78, fall80}. The
galaxy properties such as mass, morphology, colour, star formation
rate and metallicity evolve due to internal processes within galaxies
(secular evolution) and interactions with their environment. A host of
physical processes such as gas accretion, interaction and merger with
other galaxies, star formation, supernovae explosions and AGN feedback
shape the galaxy properties, leading to the diversity of galaxies in
the present universe.

The external environment play a crucial role in galaxy formation and
evolution. Understanding the roles of environment on galaxy properties
remains a complex and active area of research. Several studies
\citep{davis2, dress, butcher84, guzo, zevi, goto, hog1, blan1,
  einas2, balogh02, kauffmann04, abbas06, mouhcine, bamford, koyama}
have been carried out to understand the roles of environment on the
galaxy formation and evolution. It is now well known that morphology
\citep{dress, will, zehavi, park1}, colour \citep{baldry, pandey2,
  bamford, kreckel}, stellar mass \citep{chabrier, peng10}, star
formation rate \citep{toomre, lewis, gomez, pandey3, porter,
  kauffmann04} and many other galaxy properties are strongly
correlated with the environment. The environment is generally
characterized by the local density in most of these studies. But the
large-scale distribution of matter in the universe can not be
completely characterized in terms of density. Observations from
various redshift surveys
\citep{gregory78,joeveer78,einasto80,zeldovich82,einasto84, bharadfil,
  pandeyfil} reveal that the galaxies are distributed in a complex
interconnected network of filaments, sheets and knots surrounded by
enormous voids. This network is often referred to as the ``cosmic
web'' \citep{bond96}. The filaments are usually located at the
intersections of sheets and the galaxy clusters are located at the
intersections of filaments. The studies with N-body simulations show
that matter successively flows from voids to walls, walls to filaments
and finally from filaments onto the clusters \citep{arag10, cautun14,
  ramachandra}. It has been suggested by a number of studies with
hydrodynamical simulations that more than $80\%$ baryons in the
universe reside in filaments in the form of WHIM \citep{tuominen21,
  galarraga21}. The distribution of matter in the cosmic web can
affect the gas accretion efficiency of galaxies, influencing their
star formation rates \citep{cornu18, zhu22}. Galaxies that are located
near the centers of cosmic filaments and sheets are more likely to
receive a steady supply of cold gas from their surroundings, which can
fuel star formation and increase the mass of the galaxy. Similarly,
the galaxies near the intersections of filaments, tend to have a
higher rate of accretion of gas and a higher rate of galaxy
interactions, that can trigger intense star formation and alter their
internal structures. The galaxies located in the underdense regions of
the cosmic web tend to have a more quiescent evolution.

Some studies with hydrodynamical simulations show that the galaxy
luminosity function and mass functions depend on the cosmic web
environments \citep{metuki15, xu20}. The spin of galaxies and their
host dark matter halos also show alignment relative to cosmic web
filaments \citep{punyakoti}. Recently, \citep{das23} show that the
interacting major pairs at smaller pair separations are more star
forming in filaments compared to those hosted in sheets. These
together suggest that the geometry of the cosmic web can significantly
influence the galaxy properties and their evolution. The observational
evidence that the galaxy properties are affected by their large-scale
environment is mounting in recent years \citep{pandey2, pandey3,
  trujillo, erdogdu, paz, jones, scudder, tempel1, tempel2, darvish,
  filho, lupa, pandey17, lee18, chen, pandey20}.

The galaxy properties can also be modulated by the mass of its dark
matter halo. The intrinsic properties of the halos can influence the
galaxy properties through different physical mechanisms such as mass
quenching \citep{birnboim03, dekel06, keres05, gabor10} and angular
momentum quenching \citep{peng20}. The properties like mass, shape and
spin of the dark matter halos are sensitive to their locations inside
the cosmic web \citep{hahn1, jounghun}. Besides, the clustering of the
dark matter halos also depend on their assembly history \citep{croton,
  gao, musso, vakili}. \citep{tojeiro17} show the evidence for halo
assembly bias as a function of cosmic web environment using the GAMA
survey.

The different galaxy properties such as stellar mass, colour, star
formation rate and metallicity are known to be strongly correlated
with each other and such correlations may depend on the environment.
The blue galaxies are typically lower mass systems with ongoing star
formation, while red galaxies are more massive and have mostly ceased
star formation. The transition from blue to red galaxies is driven by
a process called quenching, in which the SFR decreases and the galaxy
becomes redder in colour. Such quenching may be driven by different
physical mechanisms such as galaxy harassment \citep{moore96,
  moore98}, starvation \citep{larson80, somerville99, kawata08},
strangulation \citep{gunn72, balogh00}, ram pressure stripping
\citep{gunn72} and gas outflows through supernovae, AGN or
shock-driven winds \citep{cox04, murray05, springel05}. A number of
internal physical processes such as morphological quenching
\citep{martig09} and bar quenching \citep{masters10} can also suppress
the star formation in galaxies. On the other hand, the star formation
activity in galaxies can be significantly enhanced due to interactions
\citep{barton00, lambas03, alonso04, nikolic04, ellison10, woods10,
  patton11, violino18, thorp22}. Both galaxy-galaxy interactions and
the interactions with the environment are crucial in shaping the
correlations between different galaxy properties. The more massive
galaxies are generally redder and found in denser environments. The
mass and SFR of galaxies are also correlated with their
metallicity. The metallicity is a measure of the abundance of elements
heavier than Helium in a galaxy. The more massive galaxies have higher
metallicities as they are able to retain their heavy elements through
repeated rounds of star formation and supernova explosions, which
eject heavy elements into the interstellar medium. The low-mass
galaxies have shallower gravitational potential wells. This makes it
easier for gas to escape during outflows which reduce the overall
metallicity of the galaxy. Observations also indicate that galaxies
with higher metallicities have lower SFRs \citep{mannucci10}.

Understanding the observed correlations between different galaxy
properties can provide us useful insights on galaxy formation and
evolution. The correlations can be qualitatively explained using the
equilibrium model of galaxy formation and evolution \citep{dekel09,
  dave11, dave12, lilly13}. It is a simple theoretical model that
assumes that the galaxies maintain a steady state, in which the inflow
of gas from the IGM is balanced by the outflow of gas due to feedback
processes such as supernova explosions and the winds generated by
AGN. Galaxies perturbed by interactions, mergers or environment are
driven back to their equilibrium and this allows the galaxies to
maintain their available gas over time. However, the equilibrium model
does not take into account all of the complex processes that
contribute to the evolution of galaxies, such as mergers,
gravitational interactions, and the influence of large-scale
structures. These can be taken into account using N-body simulations
or hydrodynamic simulations. These simulations are used to follow the
evolution of dark matter and baryonic matter in the universe, and to
study the properties of galaxies that form in different
environments. Nonetheless, our understanding of the precise nature of
the correlations and the physical mechanisms driving them are far from
being complete.

It is important to analyze the observational data to understand the
correlations between galaxy properties in different environments of
the cosmic web. The Sloan Digital Sky Survey (SDSS) \citep{stout02} is
the largest redshift survey carried out till date. It provides high
quality spectra and images of millions of galaxies in the universe
enabling us to create reliable 3D maps of the galaxy distribution and
determine the physical properties of the galaxies with unprecedented
accuracy.  In the present work, we would like to analyze the SDSS data
to study the correlations between a number of galaxy properties in
different geometric environments of the cosmic web. We intend to
investigate whether these correlations are affected by the geometry of
the cosmic web. We plan to identify the galaxies in different
environments of the cosmic web based on the signs of the eigenvalues
of the tidal tensor \cite{hahn2,fromero} at their locations. We will
analyze the correlations between stellar mass, colour, star formation
rate and metallicity of the galaxies at each type of environment. One
can calculate the Pearson correlation coefficient for each pair of
galaxy properties in different geometric environments of the cosmic
web. The Pearson correlation coefficient measures the linear
relationship between two random variables. However some of the
relationships between different galaxy properties can be non-linear
and non-monotonic. This motivates us to use the mutual information for
measuring correlation between galaxy properties. The mutual
information is a non-parametric measure for measuring correlations
between random variables and it can detect any kind of association,
including non-linear and non-monotonic relationships. We will
calculate the mutual information for each pair of galaxy properties in
different types of geometric environments and test whether they differ
in a statistically significant manner. This would help us to quantify
the roles of the cosmic web in shaping the correlations between
different galaxy properties.

The plan of the paper is as follows. We describe the data in Section
2, explain our methods in Section 3, discuss the results of our
analysis in Section 4 and finally present our conclusion in Section 5.

\section{Data}
\label{sec:data}
\begin{figure}[h!]
\includegraphics[width=7.5cm]{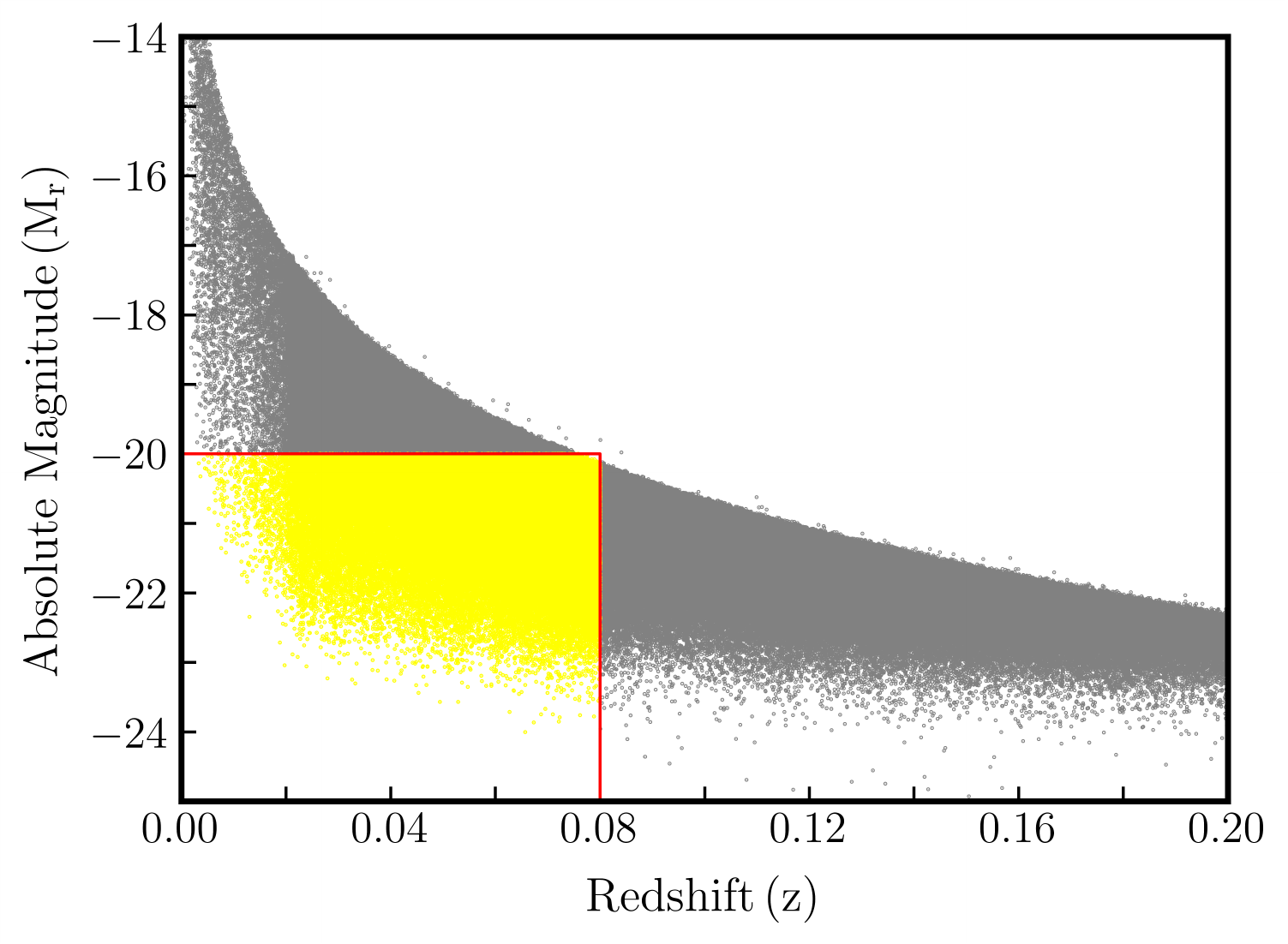}
\hspace*{0.5em}
\includegraphics[width=7.45cm]{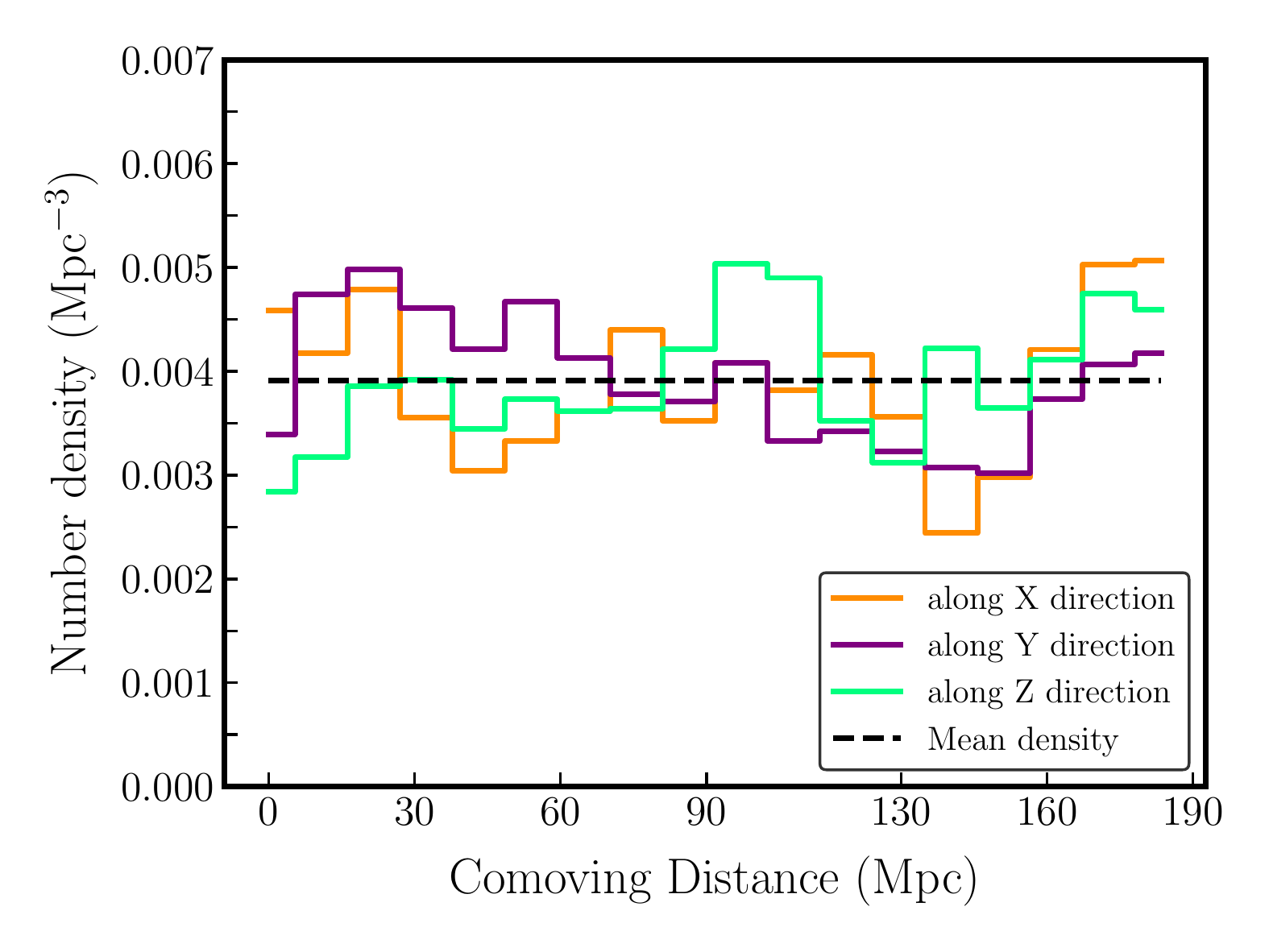}
\caption{The left panel shows the definition of the volume limited
  sample in the redshift-absolute magnitude plane. The variations of
  the comoving number density of galaxies in the extracted cubic
  region are shown in the right panel. The number density is
  calculated using slices of thickness $10 \, \rm Mpc$. }
\label{fig:1}
\end{figure}

We use the publicly available data from the seventeenth data release
\cite{abdur22} of the Sloan Digital Sky Survey (SDSS DR17) for the
present analysis. DR17 is the fifth and final data release of the
fourth phase of the SDSS survey. We use \textit{Structured Query
  Language}(SQL) to download the data from the SDSS
SkyServer \footnote{https://skyserver.sdss.org/casjobs/}. We select
all galaxies within \textit{right ascension} range $130^{\circ} \leq
\alpha \leq 230^{\circ}$ and \textit{declination} range $0^{\circ}
\leq \delta \leq 62^{\circ}$. The SDSS imposed a restriction in the
r-band Petrosian apparent magnitude to $\rm m_r \leq 17.77$, where
only photometric targets brighter than this limit were chosen for
follow-up spectroscopy. We construct a volume limited sample by
restricting the r-band absolute magnitude to $\mathrm{M_r}\,\leq
-20$. The resulting volume limited sample radially extends upto $z
\leq 0.08$ and contains a total $94986$ galaxies (\autoref{fig:1}). We
obtain the spectroscopic and photometric information of the galaxies
from the \textit{SpecObjAll} and \textit{Photoz} tables of the SDSS
database. We use the observed colour in the present analysis. The
colours are not corrected for reddening due to redshift and the
internal extinction. The stellar mass, specific star formation rate
(sSFR) and metallicity of the galaxies are derived from the table
\textit{StellarMassFSPSGranWideDust}. The sSFR is the star formation
rate per unit galaxy stellar mass. The metallicity is the mass
fraction of the elements heavier than Helium \citep{asplund09}. These
are estimated using the Flexible Stellar Population Synthesis (FSPS)
Technique \cite{conroy09}. It models the Spectral Energy Distribution
(SED) of galaxies by combining the star formation and metal enrichment
histories together with the stellar evolution and dust
attenuation. The stellar masses, metallicities and star formation
rates are constrained by comparing the observed spectroscopic and/or
photometric properties of the galaxies to the stellar population
synthesis models. The SDSS spectra are measured through a 3 arcsec
aperture which covers only a fraction of the total galaxy. This is
usually dealt by fitting the observed spectra with the stellar
population synthesis models and then applying a correction based on
the the observed colour outside of the fiber
\citep{brinchmann04}. However, the physical properties derived from
this approach contain significant systematic
uncertainties. \cite{conroy09} show that the information derived
solely from the broadband photometry are more robust.

We only consider those galaxies for which the \textit{scienceprimary}
flag is set to 1. This ensures that only the galaxies with best
spectroscopic information are included in our analysis. We also
compared the redshift distributions of the galaxies in our sample with
and without using the scienceprimary flag and find that they are
nearly identical.

We finally extract the largest cubic region within our volume limited
sample, which has a side length of $183.50\, \mathrm{Mpc}$. It
contains a total $24146$ galaxies and has a mean galaxy number density
of $\sim 3.933 \times 10^{-3} \, \mathrm{Mpc}^{-3}$
(\autoref{fig:1}). The mean intergalactic separation of the sample is
$\approx 6.34 \, \rm Mpc$.\\

\begin{figure}[h!]
\centering
\includegraphics[width = 14cm]{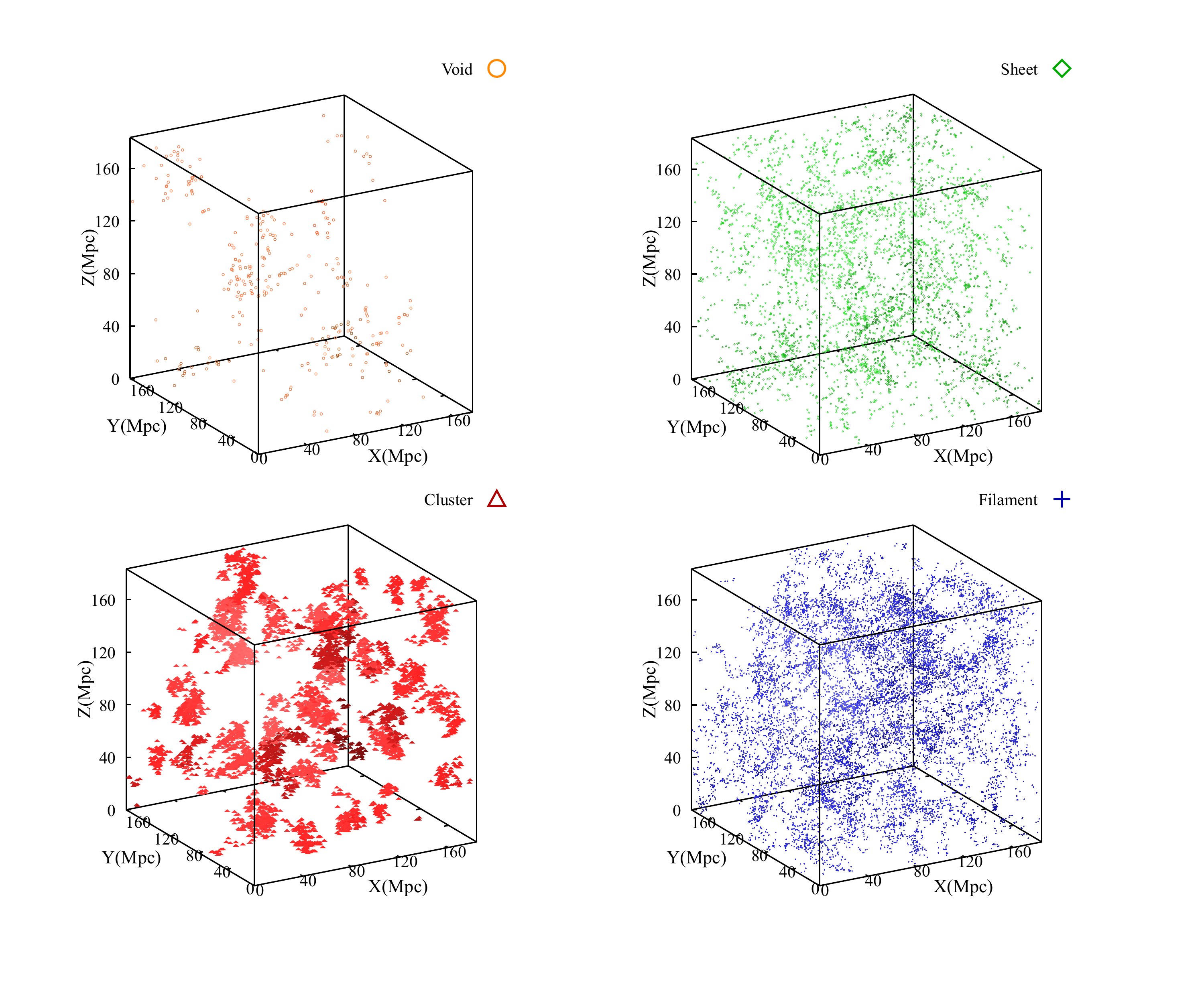}
\caption{The different panels of this figure show the
  three-dimensional distributions of galaxies in different cosmic web
  environments.}
\label{fig:2}
\end{figure}

%



\section{Method of analysis}
\subsection{Quantifying the cosmic web}
\label{sec:method}
We identify the different morphological components of the cosmic web
using the dynamical classification of the cosmic web \citep{hahn2,
  fromero}. This uses the eigenvalues of the three-dimensional tidal
tensor to classify the different morphological components. The tidal
tensor $T_{\alpha \beta}$ in 3D is defined as the Hessian of the
gravitational potential $\Psi$,
\begin{equation}
T_{\alpha \beta} = \frac{\partial^2 \Psi}{\partial x_{\alpha}\partial x_{\beta}}
\end{equation}
We first solve the Poisson equation
\begin{equation}
\nabla^2 \Psi \equiv \delta
\end{equation}
to obtain the potential. Here $\delta =
\frac{\rho-\bar{\rho}}{\bar{\rho}}$ is the density contrast. The
galaxy distribution is first converted to a density field by applying
the Cloud-in-Cell (CIC) scheme on $(128)^3$ grid with a grid spacing
of $1.43\, \rm Mpc$. We transform the resulting density field into the
Fourier space and multiply it with a Gaussian window function. We use
a smoothing length of $7.15\, \rm Mpc$ to achieve a smoothed density
distribution. Our galaxy sample does not have a high density. The
intergalactic separation in our sample is $\sim 6.34\, \rm Mpc$. The
chosen smoothing scale is 5 times the grid spacing, and is close to
the intergalactic separation. This choice limits our ability to
characterize the environment on scales smaller than the intergalactic
separation. In the present work, we are primarily interested in
quantifying the large-scale geometric environments in the cosmic web.

We obtain the Fourier transform of the gravitational potential
corresponding to the fluctuations in the smoothed density field,
\begin{equation}
\hat{\Psi} = \hat{\mathcal{G}}\hat{\rho}
\end{equation}
where $\hat{\mathcal{G}}$ is the Fourier counterpart of the Green's
function of the Laplacian operator and $\hat{\rho}$ is the density in
Fourier space. We inverse transform the potential back into the real
space and calculate the tidal tensor using numerical differentiation
of the potential. We classify each galaxy to be part of a void, sheet,
filament or cluster based on the signs of the three eigenvalues
($\lambda_1$, $\lambda_2$, $\lambda_3$) of the tidal tensor (
\autoref{tab:1}).
\begin{table}[h!]
\centering
\begin{tabular}{|c|c|c|c|}
\hline
{Morphological environment} & {$\lambda_1$} & {$\lambda_2$} & {$\lambda_3$} \\
\hline
{Void} & {$<0$} & {$<0$} & {$<0$} \\
\hline
{Sheet} & {$>0$} &{$<0$} & {$<0$} \\
\hline
{Filament} & {$>0$} &{$>0$} &{$<0$} \\
\hline
{Cluster} & {$>0$} & {$>0$} & {$>0$} \\
\hline
\end{tabular}
\caption{This table shows the signs of the three eigen values for each
  type of morphological environment.}
\label{tab:1}
\end{table}

\begin{table}[h!]
\centering
\begin{tabular}{|c|c|c|}
\hline
{Morphological environment} & {Number of galaxies present} & {Number of mass-matched galaxies} \\
\hline
{Void} & {$324$} & {$0$} \\
{Sheet} & {$4435$} & {$4363$} \\
{Filament} & {$12620$} & {$4363$} \\
{Cluster} & {$6767$} & {$4363$} \\
\hline
\end{tabular}
\caption{This table shows the number of galaxies identified in
  different morphological environments, using the criteria listed in
  \autoref{tab:1}.}
\label{tab:2}
\end{table}

The number of galaxies identified in different morphological
environments are listed in \autoref{tab:2}. We show the
three-dimensional distributions of the galaxies identified in
different types of geometrical environments in different panels of
\autoref{fig:2}. Our primary aim in this study is to compare the
influence of geometric environments on the correlations between
different galaxy properties. We do not consider the void galaxies in
our analysis due to their lower abundance.


\begin{figure}[h!]
\hspace{-20px}
\includegraphics[width = 16cm]{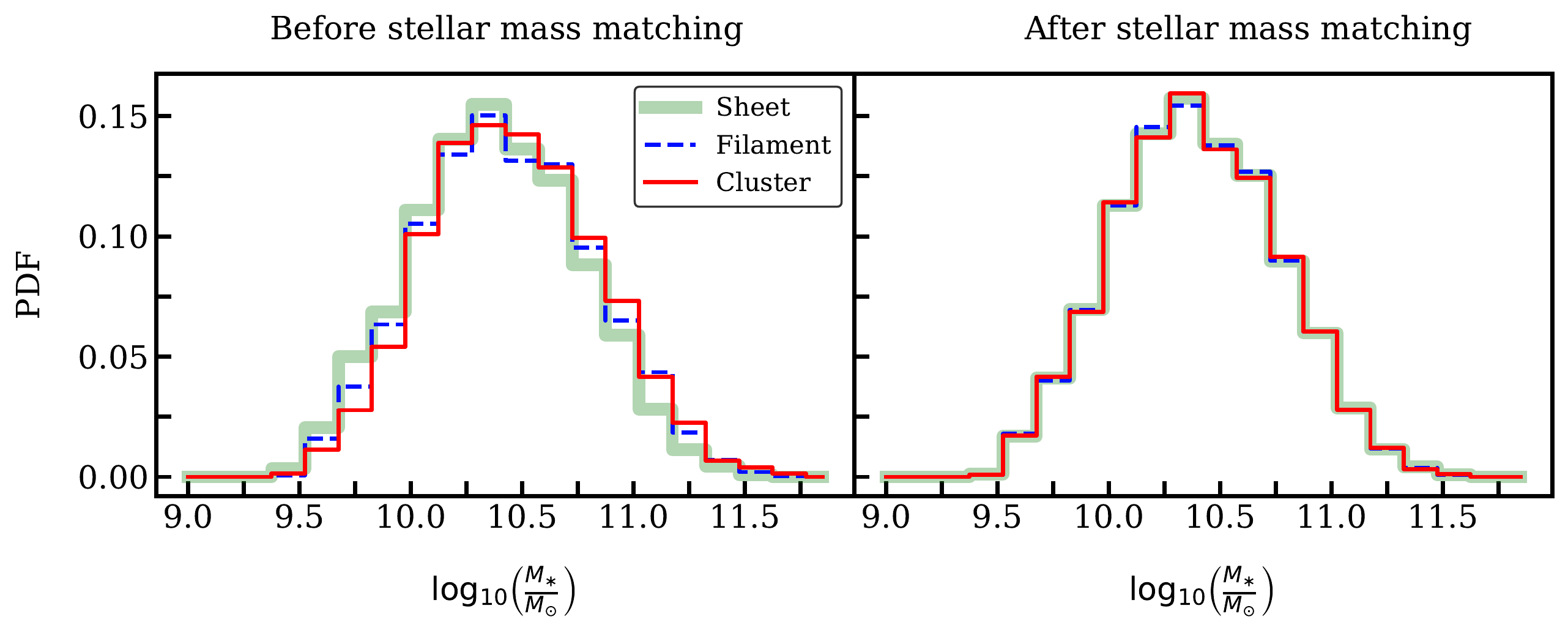}
\caption{The left panel and right panel of this figure respectively shows the PDF of the stellar mass distributions in different environments before and after matching the stellar mass.}
\label{fig:matching}
\end{figure}


The mass of a galaxy is one of the most important factors influencing
the galaxy properties. The strong correlations with the stellar mass
can seriously bias our study unless we properly take this into
account. The differences in the correlations across various
environments may also arise due to the difference in the stellar mass
distributions in these environments. It is important to ensure that
the results of these analysis are not affected by the link between
stellar mass and environment. The sheetlike environments host 4435
galaxies in our dataset. We would like to draw the same number of
galaxies from filaments and clusters after matching their mass
distributions.This is done in order to have same shot noise
contributions in our measurements at each environment. We match the
stellar mass (within $0.01$ dex) of equal number of galaxies from the
three geometric environments. This provides us with 4363 galaxies in
each environment (\autoref{tab:2}). We then use a Kolmogorov-Smirnov
(KS) test to compare the PDFs of the stellar mass distribution in
different environments. The left and right panels of
\autoref{fig:matching} respectively show the PDFs before and after
stellar mass matching. We find that the null hypothesis can be
accepted at a very high confidence level ($>99.99\%$) for the stellar
mass distributions at each pair of environment. This ensures that the
stellar mass distributions of the galaxies in the three different
geometric environments are statistically indistinguishable from each
other. We carry out our entire analysis using the stellar mass matched
samples.

It is worthwhile to mention some limitations of our analysis. The
present analysis is carried out in the redshift space, where the
“Finger of God” (FOG) effect in the high density groups and clusters
may introduce some spurious filaments. Our analysis does not take this
effect into account. The volume limited sample used in our analysis
does not contain many high density groups/clusters due to the fibre
collision effects. Nonetheless, the SDSS spectroscopic sample have
completeness issues due to the inherrent limitations from the fibre
collisions. The incompleteness issues should be taken into account for
a reliable estimation of the environment on small scales. In this
work, we are primarily concerned about the large-scale cosmic web
environments. These issues may have some impact on our
results. However, we do not expect these issues to bias our results in
a serious manner.

\subsection{Measuring the correlation between different galaxy properties}
 
We use Pearson correlation coefficient and normalized mutual
information (NMI) for measuring the correlations between galaxy
properties in different geometric environments of the cosmic web. 

\subsubsection{Pearson correlation coefficient}
The Pearson correlation coefficient (PCC) is the simplest measure of
linear association between two random variables. It provides the
magnitude of the correlation as well as the direction of the
relationship.

The PCC for a pair of random variables ($X,Y$) is defined as,
\begin{equation}
r_{XY} = \frac{\sum_{i=1}^{N} (X_i-\bar{X})(Y_i-\bar{Y})}{\sqrt{\sum_{i=1}^{N}(X_i-\bar{X})^2 \sum_{i=1}^{N}(Y_i-\bar{Y})^2}}
\end{equation}
where $\bar{X} = \frac{1}{N}\sum_{i=1}^{N}X_i$ and $\bar{Y} =
\frac{1}{N}\sum_{i=1}^{N}Y_i$ are the average values of $X$ and $Y$
and $N$ is the total number of data points. The PCC ranges from -1 to
1. The values -1 and 1 respectively convey perfect negative
correlation and perfect positive correlation between the two random
variables.

\subsubsection{Normalized mutual information}
The relationship between some galaxy properties can be also non-linear
and non-monotonic. The mutual information is an information theoretic
measure that can detect any kind of association, including non-linear
and non-monotonic relationships. Further, it is a non-parametric
measure and does not assume a specific distribution for the variables,
whereas Pearson correlation assumes a normal distribution. Therefore,
the mutual information can be considered to be a more general and
robust measure of association compared to the Pearson correlation
coefficient.

If $X$ is a discrete random variable which has $n$ possible outcomes
\{$X_i : i = 1,\ldots n$\} and $P(X_i)$ is the probability for the
$i^{th}$ outcome, then the Shannon entropy corresponding to $X$ is
defined as,
\begin{equation}
H(X) = -\sum_{i=1}^{n} P(X_i)\,\log P(X_i).
\end{equation} 
where the base of the logarithm is chosen to be $10$.  We consider two
discrete random variables $X$ and $Y$ that represent two distinct
galaxy properties. The joint entropy of the two variables is
determined as,
\begin{equation}
H(X,Y) = -\sum_{i=1}^{n_1}\sum_{j=1}^{n_2} P(X_i,Y_j)\,\log P(X_i,Y_j).
\end{equation}
where $P(X,Y)$ is the joint probability distribution of $X$ and $Y$.
We choose the number of bins to be $ n_1 = n_2 = 20$ for our analysis.

The mutual information (MI) between $X$ and $Y$ is defined as,
\begin{equation}
I(X;Y) = H(X) + H(Y) - H(X,Y)
\end{equation}
The MI provides a measure of the amount of information one variable
conveys about another, regardless of the relationship between them.
It does not assume a linear relationship between the two
variables. The MI is zero when the two variables are statistically
independent.

The NMI \citep{strehl02} is a normalized version of the mutual
information. It is defined as,
\begin{equation}
NMI (X;Y) = \frac{I(X;Y)}{\sqrt{H(X)H(Y)}}
\end{equation}
NMI ranges between 0 and 1, with 1 indicating strongest association
between the variables and 0 indicating no association.

The entropy is known to be sensitive to the binning. However this
should not be a problem as long as the comparisons are carried out
using the same number of bins. We use same number of bins and same
number of galaxies for the analysis in each type of geometric
environment. This would ensure same level of contributions from shot
noise at each environment and allow us to compare the results across
the environments in a meaningful way.

We also consider the effects of the binning by repeating our NMI
analysis for different number of bins (\autoref{sec:appen}).

\section{Results}

\begin{figure}[h!]
\includegraphics[width=8cm]{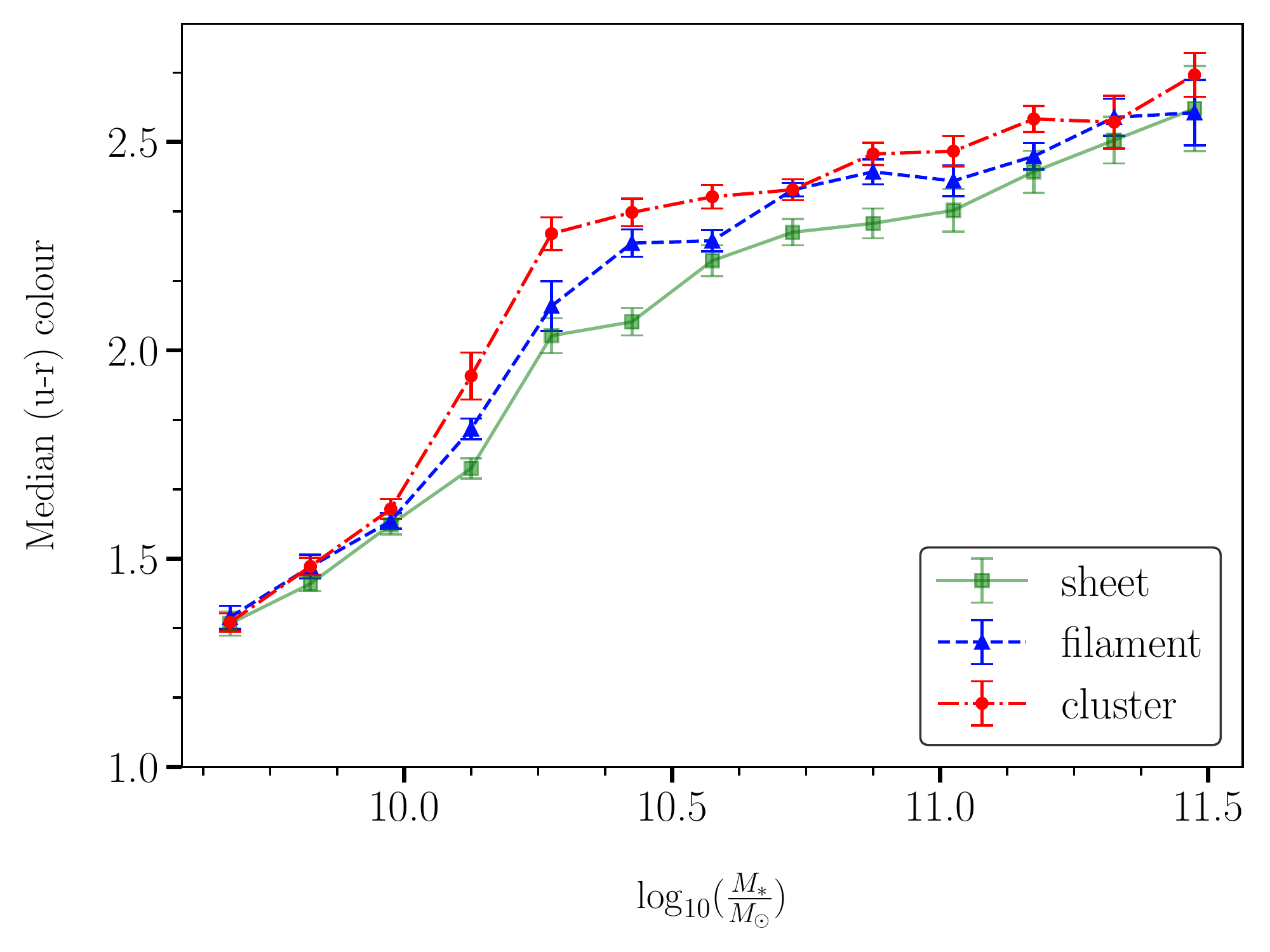}
\includegraphics[width=8cm]{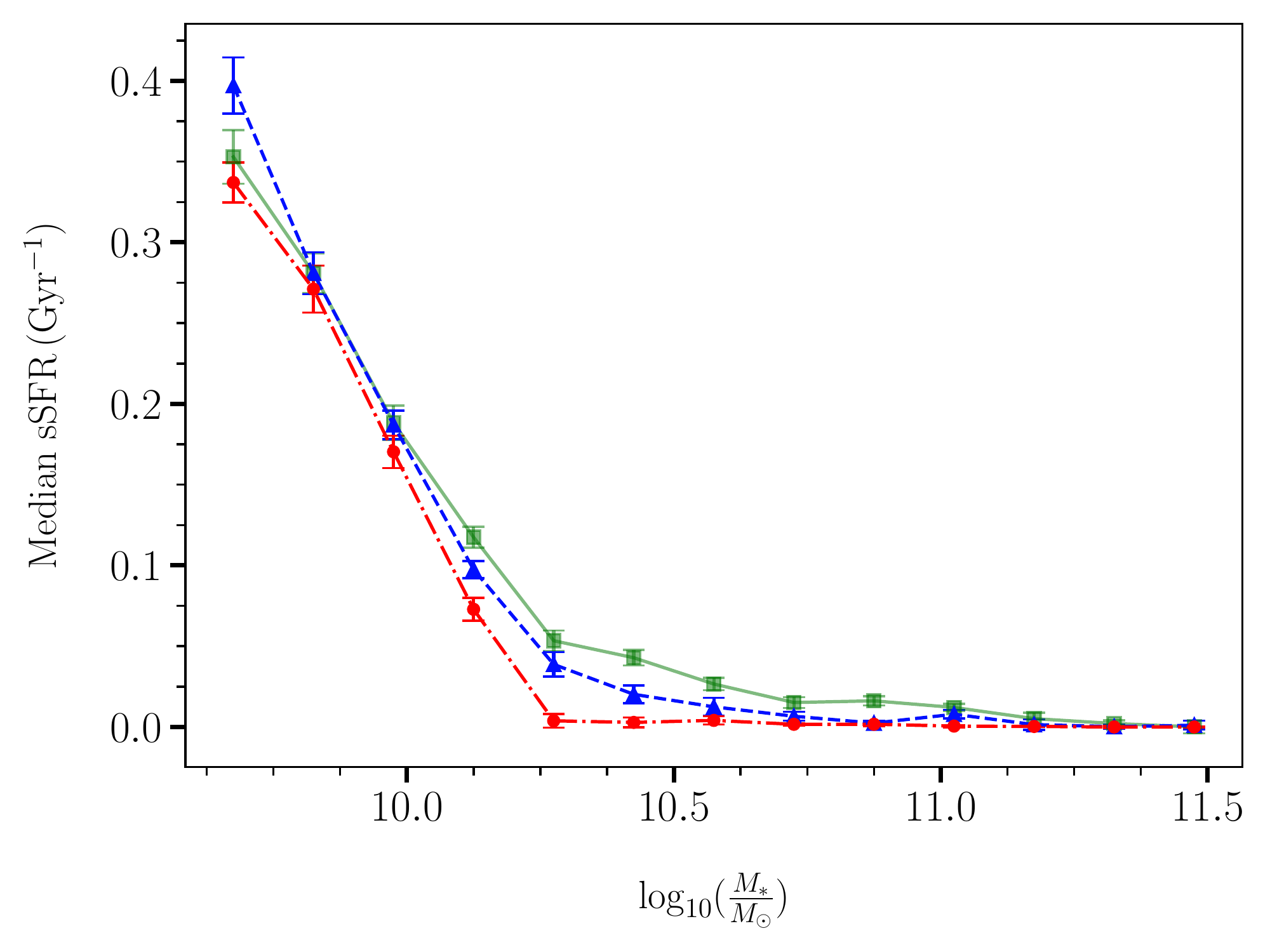}
\includegraphics[width=8cm]{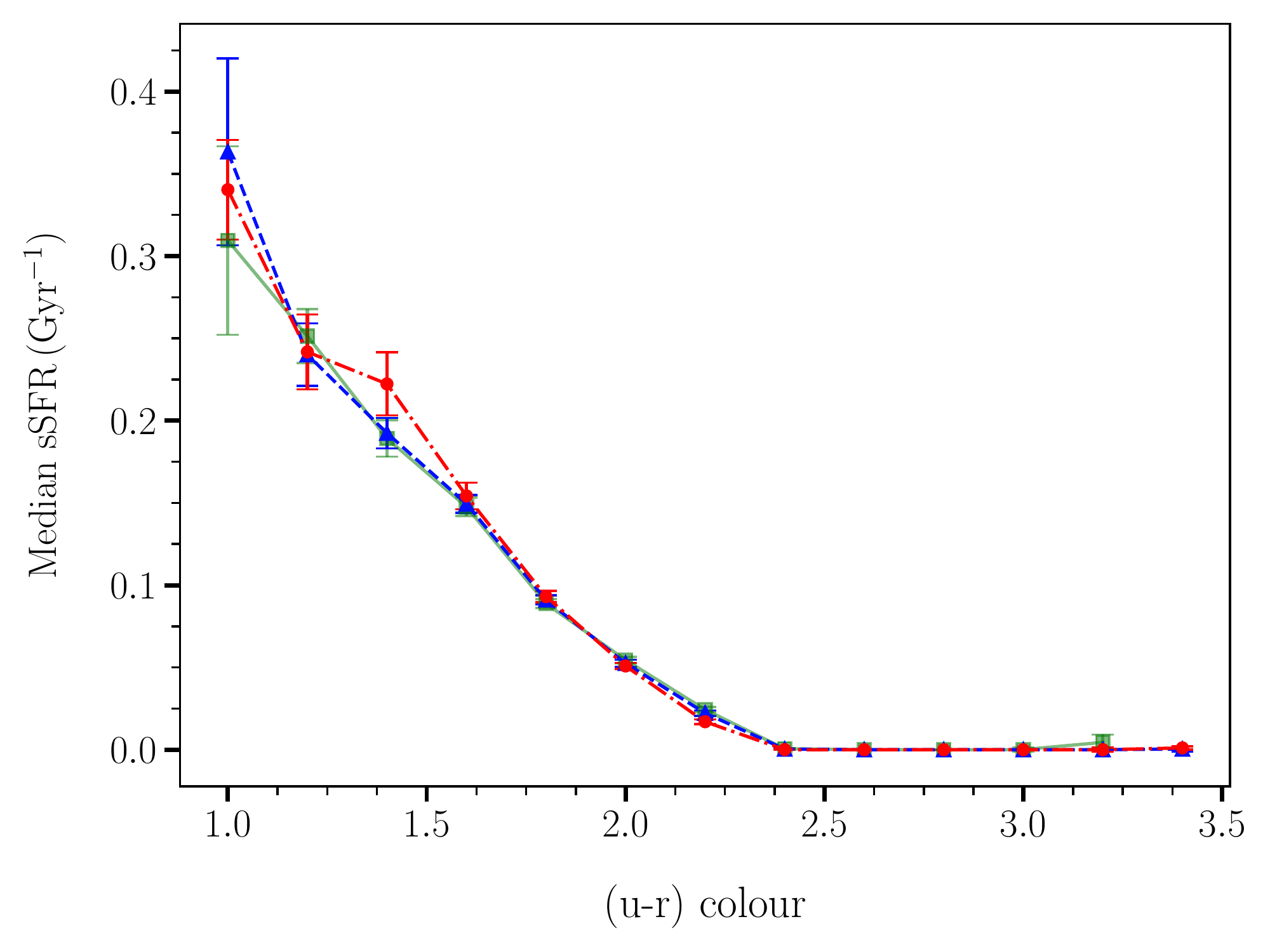}
\includegraphics[width=8cm]{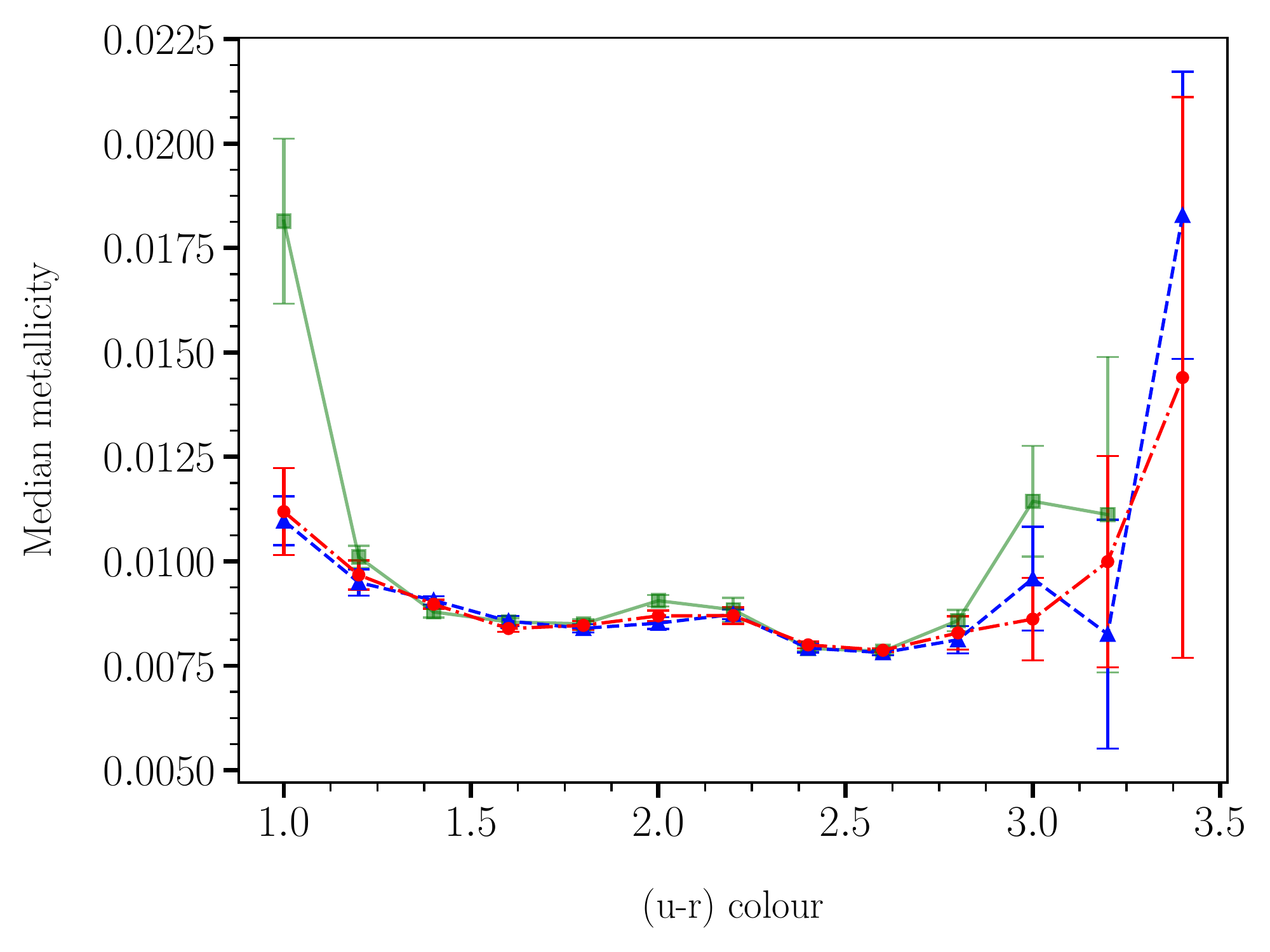}
\includegraphics[width=8cm]{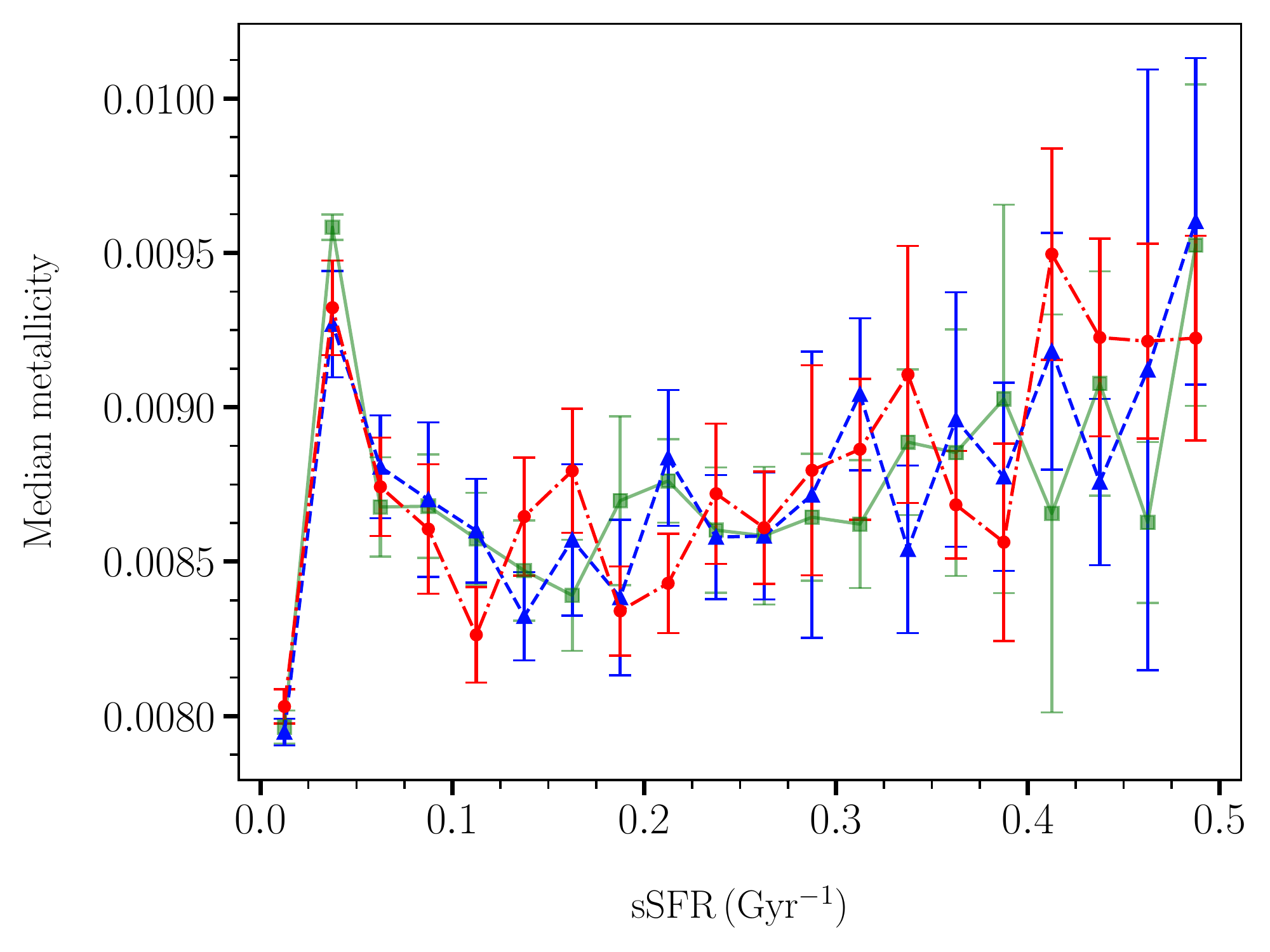}
\hspace{0.8cm}
\includegraphics[width=8cm]{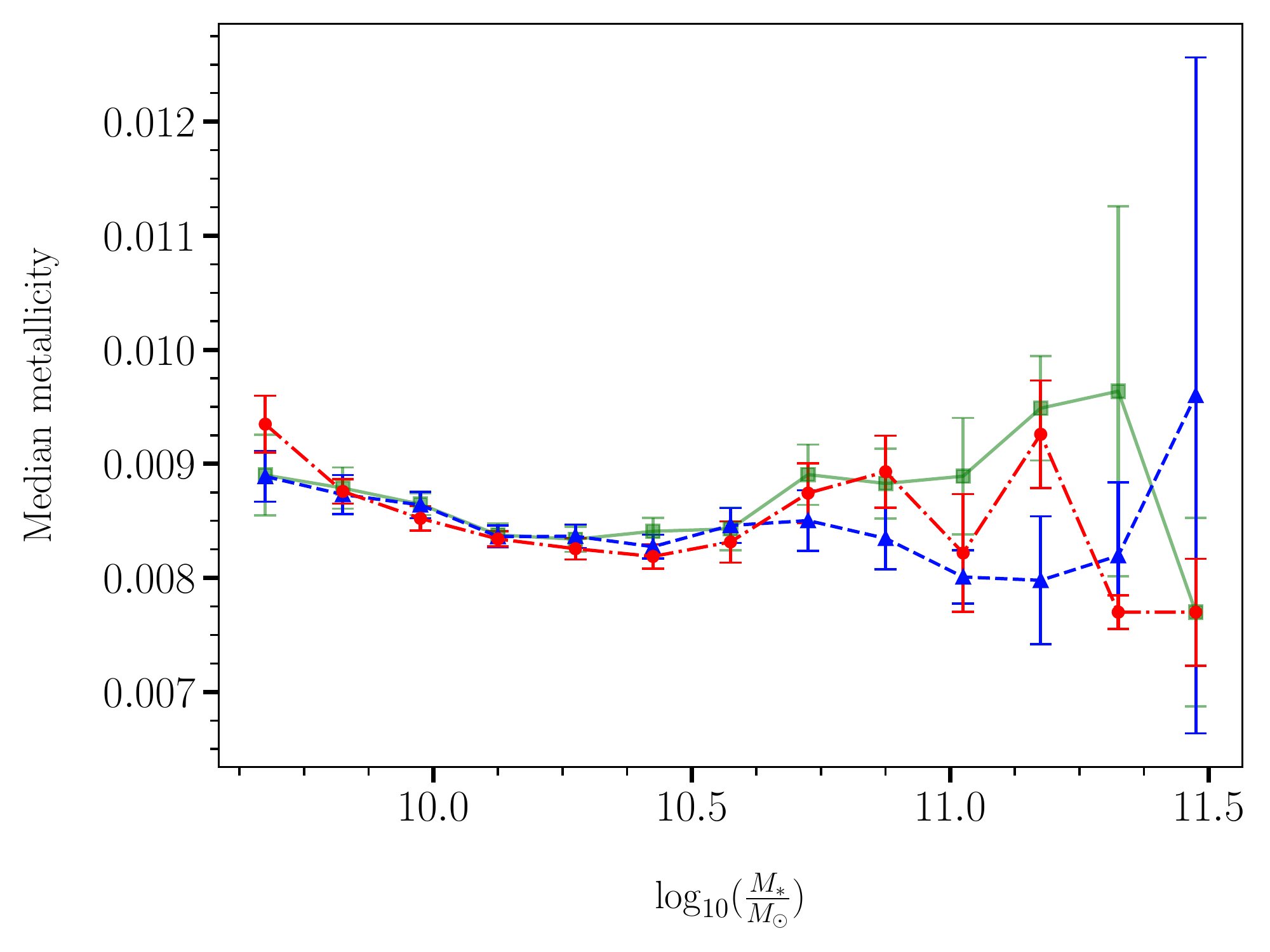}
\caption{Different panels of this figure show the relation between
  different pairs of galaxy properties. The x and y axes in each panel
  represent two different galaxy properties. We do not show the
  scatter plots for clarity and only show the median values along the
  y-axes for each values of x. The results for the galaxies in
  filaments, sheets and clusters are shown together in each panel for
  comparison. The 1$\sigma$ errors shown on each data point are
  obtained from 50 bootstrap samples drawn from the original
  datasets.}
\label{fig:3}
\end{figure}

\begin{figure}[htbp!]
\centering
\includegraphics[width=15cm]{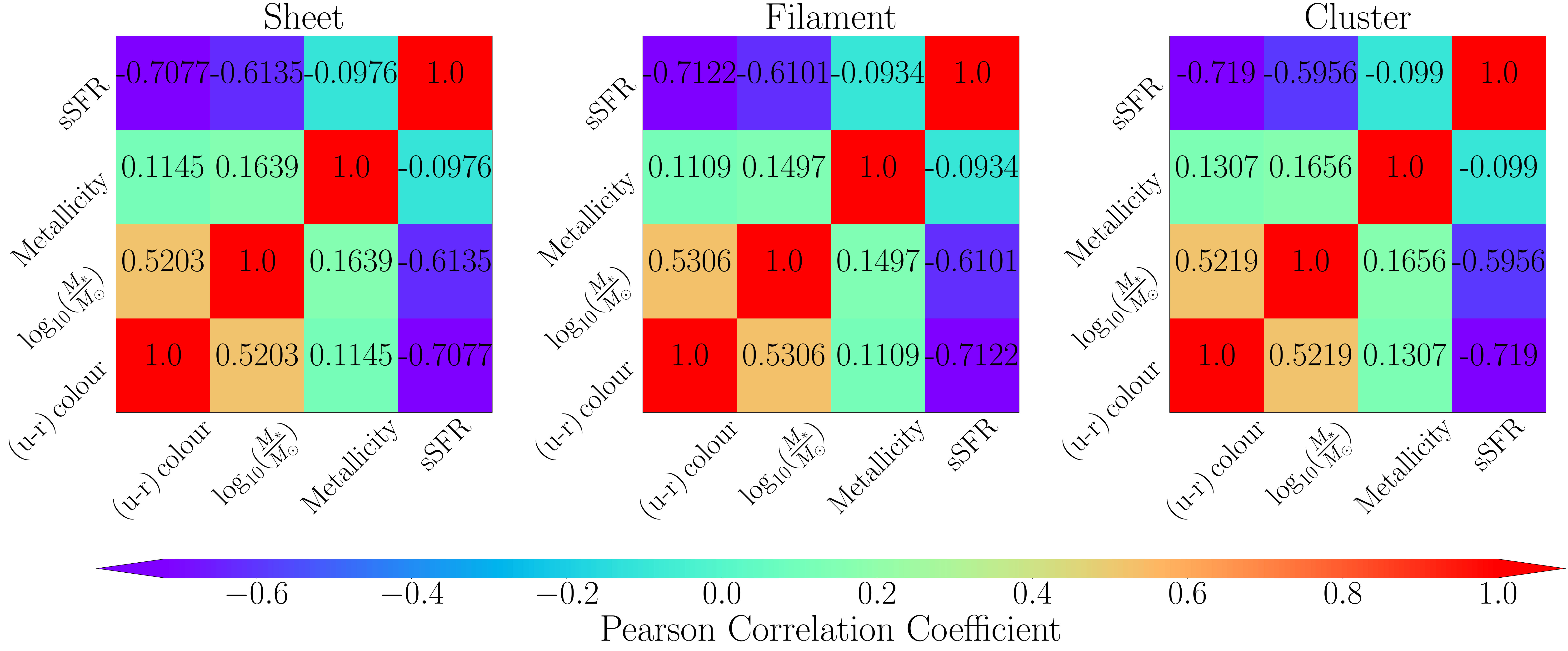}
\vspace{20px}
\centering
\includegraphics[width = 15cm]{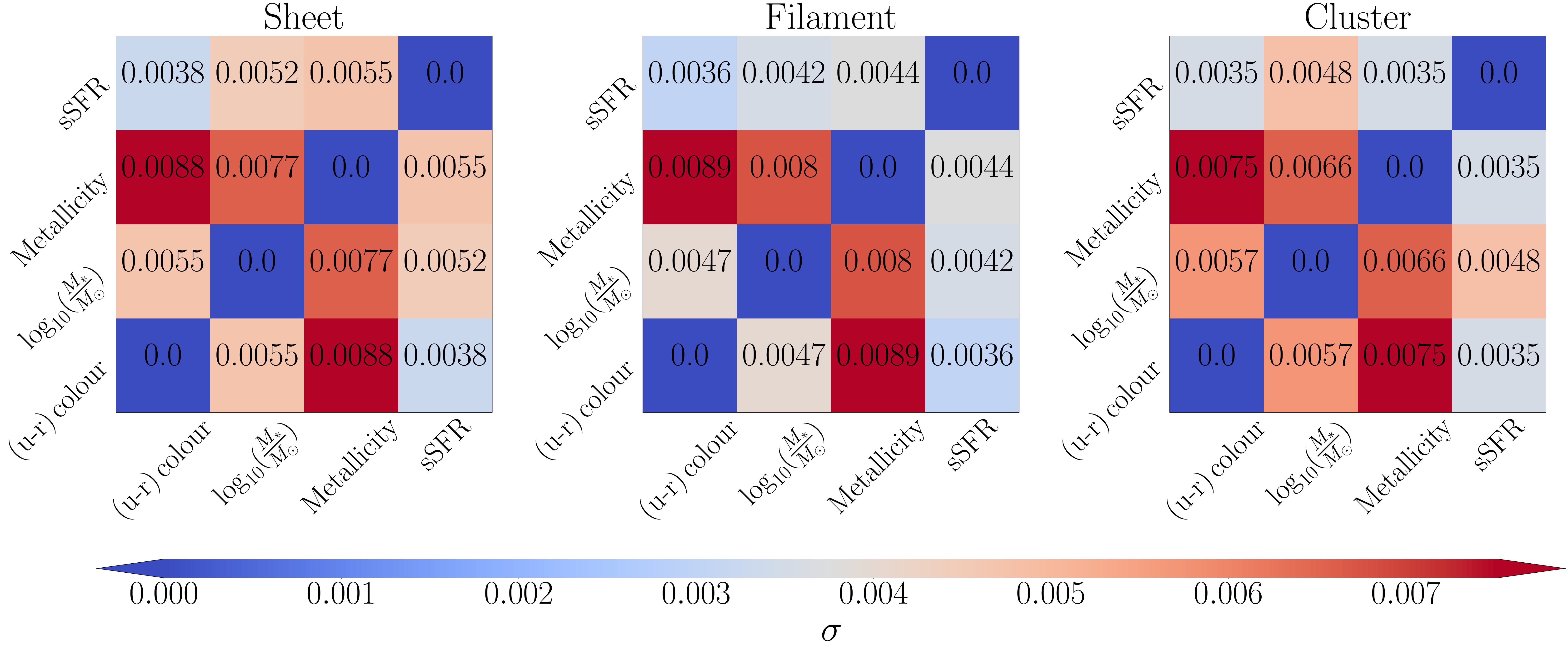}
\caption{The top three panels of this plot show the Pearson
  correlation coefficients (PCC) for each pair of the four galaxy
  properties, $\rm (u-r)\,colour,\,
  \log_{10}(\frac{M_{\ast}}{M_{\odot}}), \, metallicity, \rm sSFR$ in
  sheet, filament and cluster, after mass matching. The $1\sigma$
  standard deviations corresponding to these measurements are shown in
  the three bottom panels. The error bars in each case are obtained
  from 50 jackknife samples drawn from the original datasets.}
\label{fig:4}
\end{figure}

\subsection{Correlations between galaxy properties in filaments, sheets and clusters}
\label{sec:median}
We study the correlations between colour, stellar mass, sSFR and
metallicity of galaxies in different morphological environments. One
can use the scatter plots to investigate the relations between
different galaxy properties. However it would be difficult to make a
quantitative assessment of the correlations and compare them across
different environments. Some of these relations may not be linear and
monotonic. We consider a pair of galaxy properties and divide one of
them into a number of equal sized bins. We estimate the median value
of the other property in each of these bins. We show the binned galaxy
property along the x-axes and the median value of the other galaxy
property in each bin along the y-axes in different panels of
\autoref{fig:3}. The results for the galaxies in filaments, sheets and
clusters are shown together in each panel for comparison.

In the top left panel of \autoref{fig:3}, we find that the median
colour of galaxies in each type of morphological environment increases
with the increasing stellar mass. The enhancement of median colour
with stellar mass is strongest in the cluster environment followed by
the filamentary and sheet-like environments. The differences are more
visible above a stellar mass of $10^{10} \, M_{\odot}$. We note that
there is a distinct break at $\sim 3 \times 10^{10} \, M_{\odot}$
above which the slope of the curves become shallower at each
environment.  We examine the relationship between the stellar mass and
sSFR in the top right panel of \autoref{fig:3}. We find that the
median sSFR decreases with increasing stellar mass. This indicates
that star formation plays a more important role in the growth of
low-mass galaxies than high-mass galaxies at each environment. A
distinct break is also clearly visible at $\sim 3 \times 10^{10} \,
M_{\odot}$ in every environment. The galaxies above this mass appears
to be very low star forming and redder. Thus the colour and sSFR
respectively shows a correlation and an anti-correlation with the
stellar mass of galaxies. This is consistent with the findings of
\citep{kauffmann03} that the galaxies with stellar mass below $3
\times 10^{10}\,M_{\odot}$ are actively star forming whereas those
having a larger mass are mostly quenched.

The correlation between colour and stellar mass and an anti-correlation
between sSFR and stellar mass suggest that colour and sSFR should be
anti-correlated.  In the middle left panel of \autoref{fig:3}, we note
that the sSFR and colour of galaxies are anti-correlated in all
environments. It may be noted that the galaxies with $(u-r)$ colour
above $\sim 2.3$ are very low star forming. These galaxies are expected
to be part of the red sequence \citep{strateva01, pandey20a}.

We examine the relation between the $(u-r)$ colour and metallicity in
the middle right panel of \autoref{fig:3}. This indicates that the
galaxies with redder colour have a higher metallicity in each type of
morphological environment. This behaviour is more prominent in the
cluster and filament type environments. On the other hand, the bluer
galaxies tend to be metal rich in sheet-like environment.

The bottom left panel of \autoref{fig:3} show the median metallicity
of galaxies as a function of sSFR in each type of morphological
environment. This indicates a weak correlation between these two
galaxy properties in every environment. We show the median metallicity
of galaxies as a function of their stellar mass in different
environments in the bottom right panel of \autoref{fig:3}. The results
suggest that the massive galaxies tend to be metal rich in all
environments.

\begin{table}
\centering
\begin{tabular}{|c|c|c|c|c|c|c|}
\hline
{\rule{0pt}{4ex} Relations}& \multicolumn{2}{|c|}{ Sheet - Filament } & \multicolumn{2}{c|}{Filament - Cluster} & \multicolumn{2}{c|}{ Sheet-Cluster } \\ \cline{2-7} 
& $t$ score & $p$ value & $t$ score & $p$ value & $t$ score & $p$ value \\
\hline\hline
\rule{0pt}{3ex} colour-stellar mass & $-8.84$ & $3.99 \times 10^{-14}$ & $8.95$ & $2.25 \times 10^{-14}$ & $-1.41$ & $1.63 \times 10^{-1}$ \\
\hline
\rule{0pt}{3ex} colour-metallicity & $1.33$ & $1.85 \times 10^{-1}$ & $-11.95$ & $ 7.54 \times 10^{-21}$ & $-10.41$ & $1.58 \times 10^{-17}$ \\
\hline
\rule{0pt}{3ex} colour-sSFR & $6.93$ & $4.39 \times 10^{-10}$ & $10.30$ & $2.71 \times 10^{-17}$ & $15.07$ & $2.88 \times 10^{-27}$ \\
\hline
\rule{0pt}{3ex} stellar mass-metallicity  & $11.17$ & $3.54 \times 10^{-19}$ & $-11.85$ & $1.23 \times 10^{-20}$ & $-0.39$ & $6.98 \times 10^{-1}$ \\
\hline
\rule{0pt}{3ex} stellar mass-sSFR & $-3.29$ & $1.39 \times 10^{-3}$ & $-21.94$ & $1.36 \times 10^{-39}$ & $-23.85$ & $1.34 \times 10^{-42}$ \\
\hline
\rule{0pt}{3ex} sSFR-metallicity & $-4.27$ & $4.54 \times 10^{-5}$ & $5.58$ & $2.18 \times 10^{-7}$ & $0.55$ & $5.87 \times 10^{-1}$\\
\hline
\end{tabular}
\caption{This table summarizes the results of a two-tailed t-test
  after comparing the Pearson correlation coefficients for different
  pairs of galaxy properties in two different cosmic web
  environments. The degrees of freedom in this test is $98$.}
\label{tab:3}
\end{table}

\begin{figure}[h!]
\includegraphics[width=0.5\textwidth]{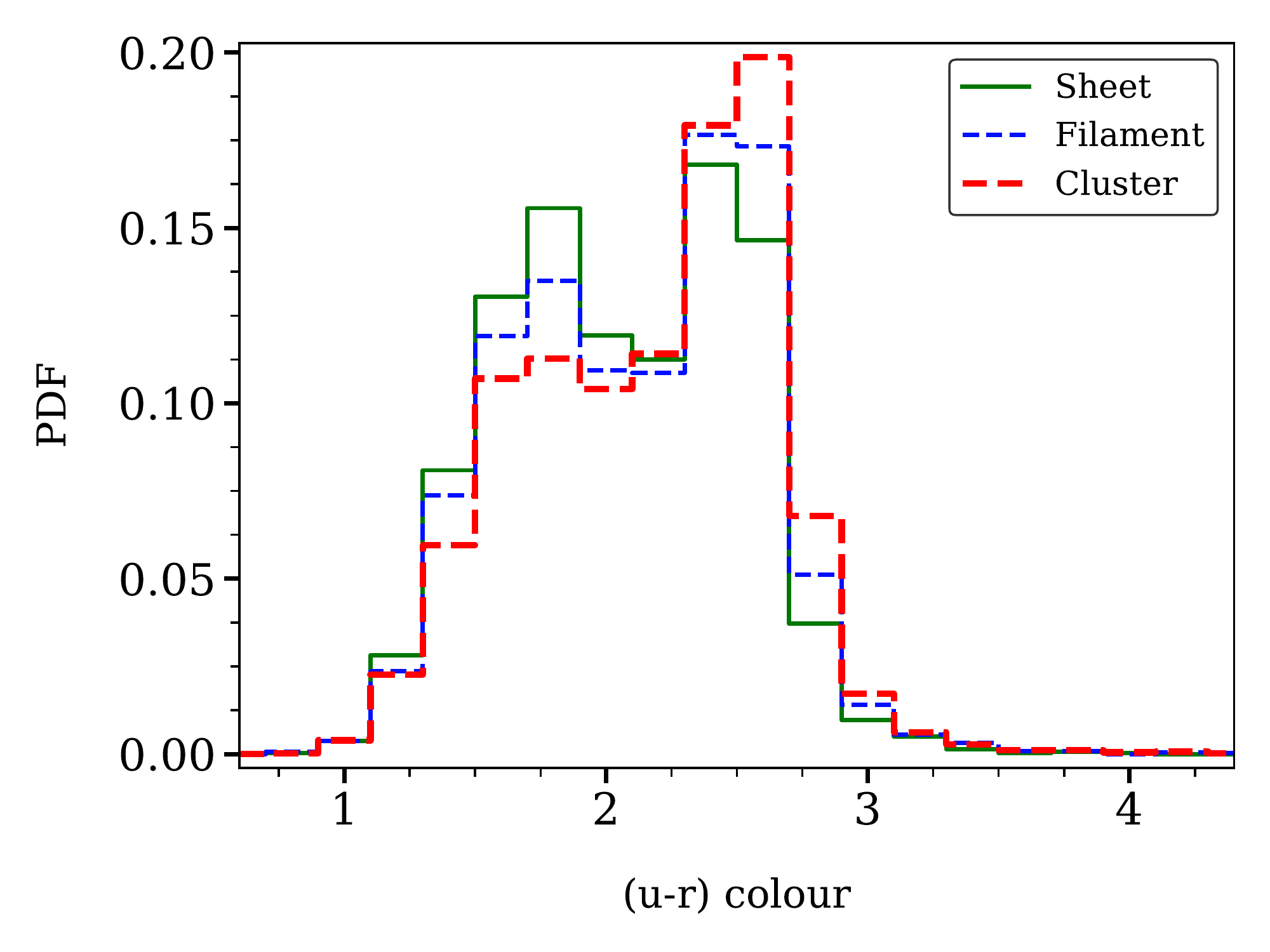}
\includegraphics[width=0.5\textwidth]{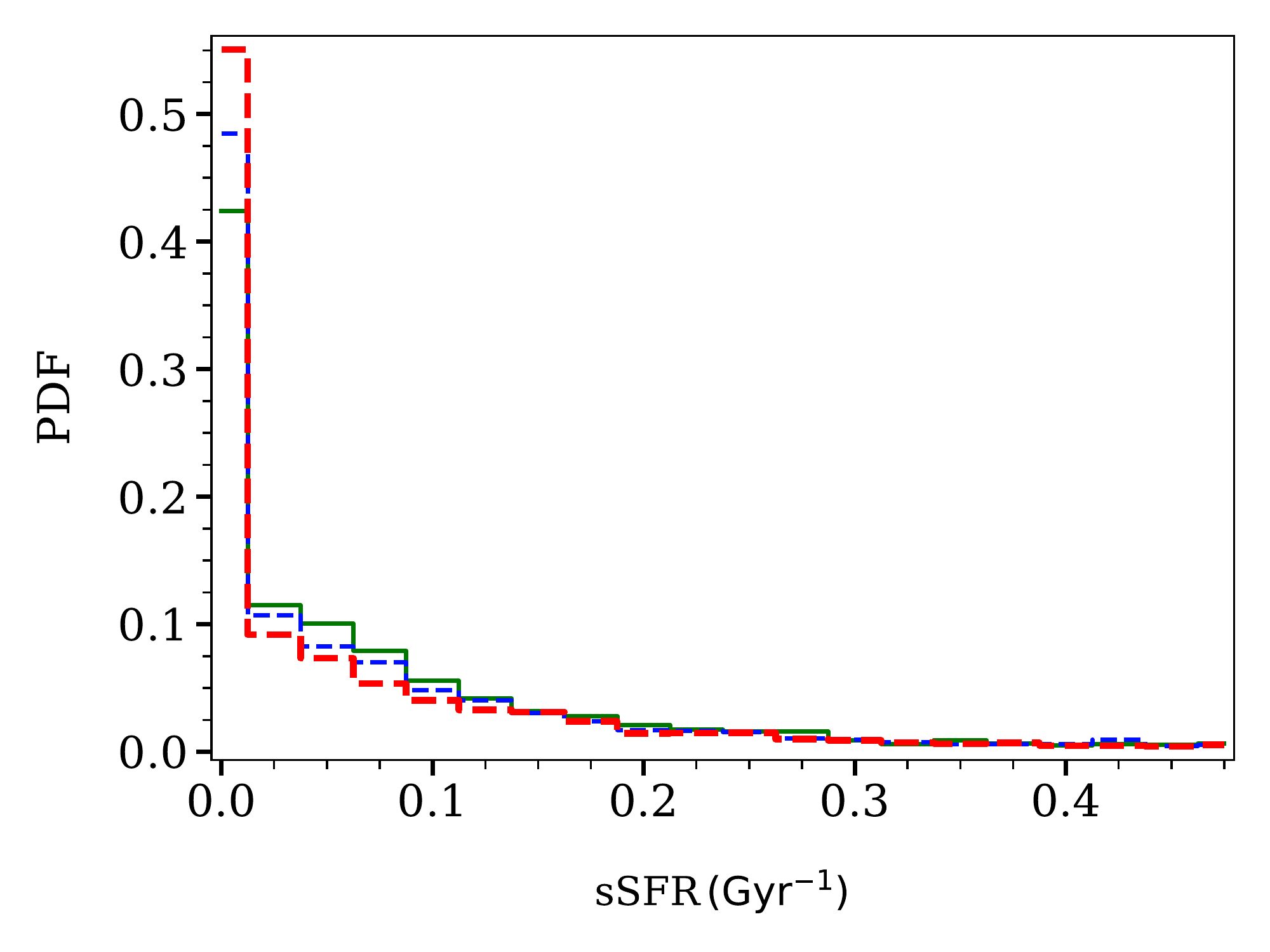}
\hspace*{4cm}
\includegraphics[width=0.5\textwidth]{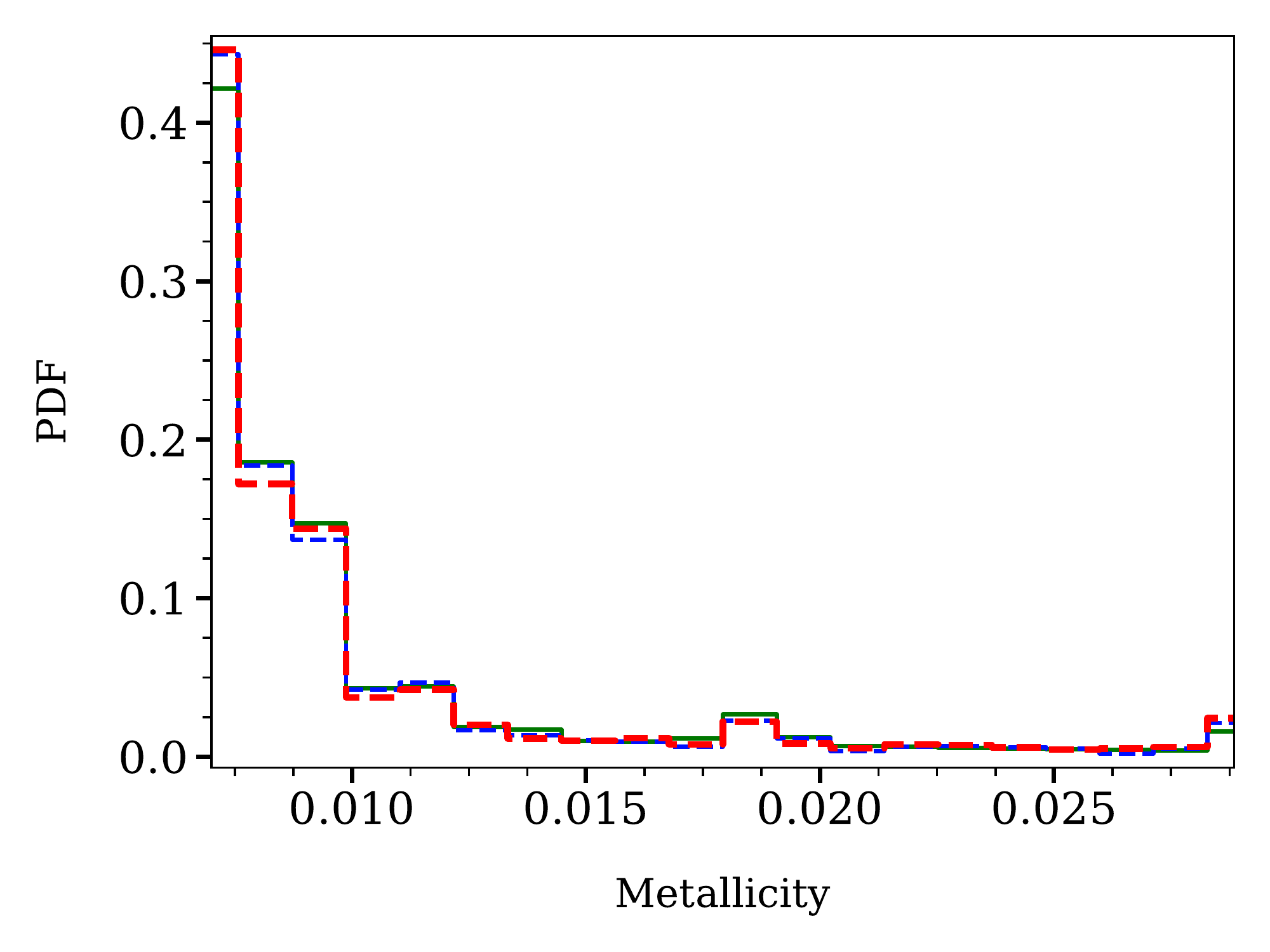}
\caption{The different panels of this figure show the probability
  distribution functions (PDF) of $\rm (u-r)$ colour, $\rm sSFR$, and
  metallicity. The PDFs in sheets, filaments and clusters are shown
  together in each panel for comparison.}
\label{fig:5}
\end{figure}

\begin{figure}
\includegraphics[width=16cm]{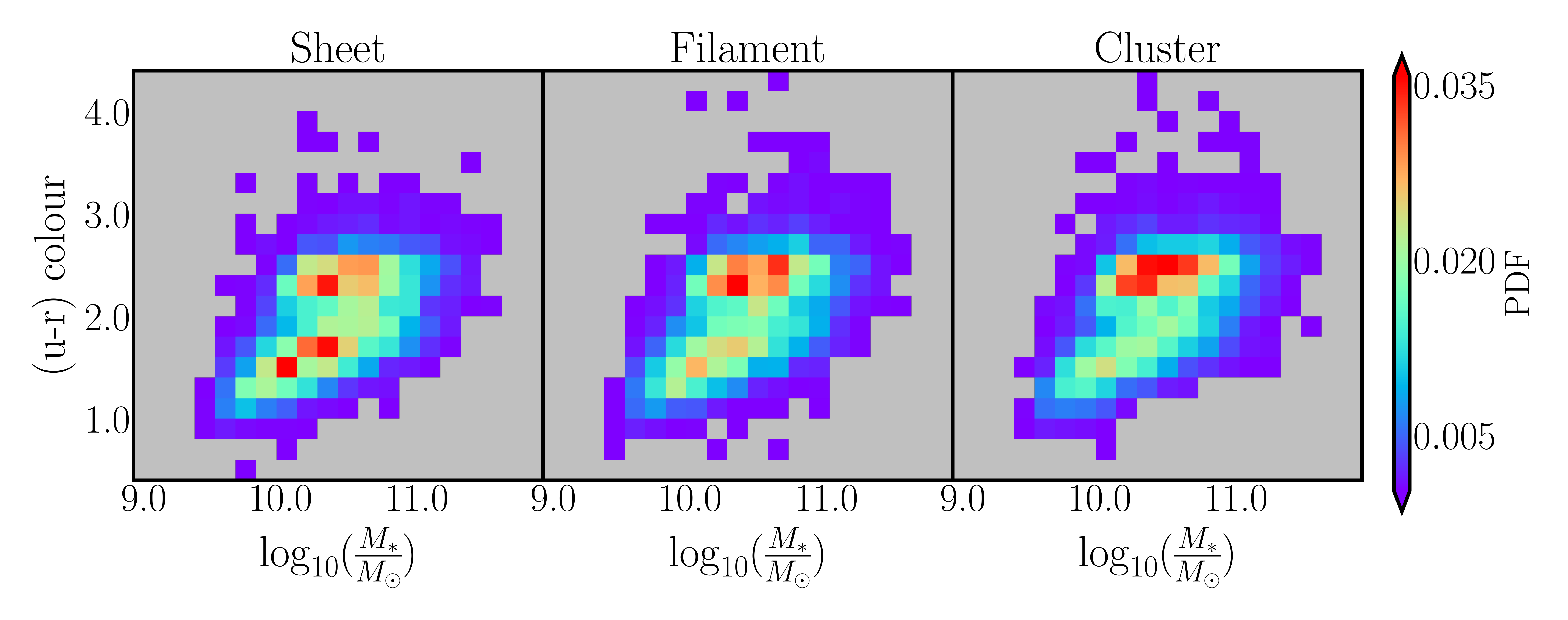}
\includegraphics[width=16cm]{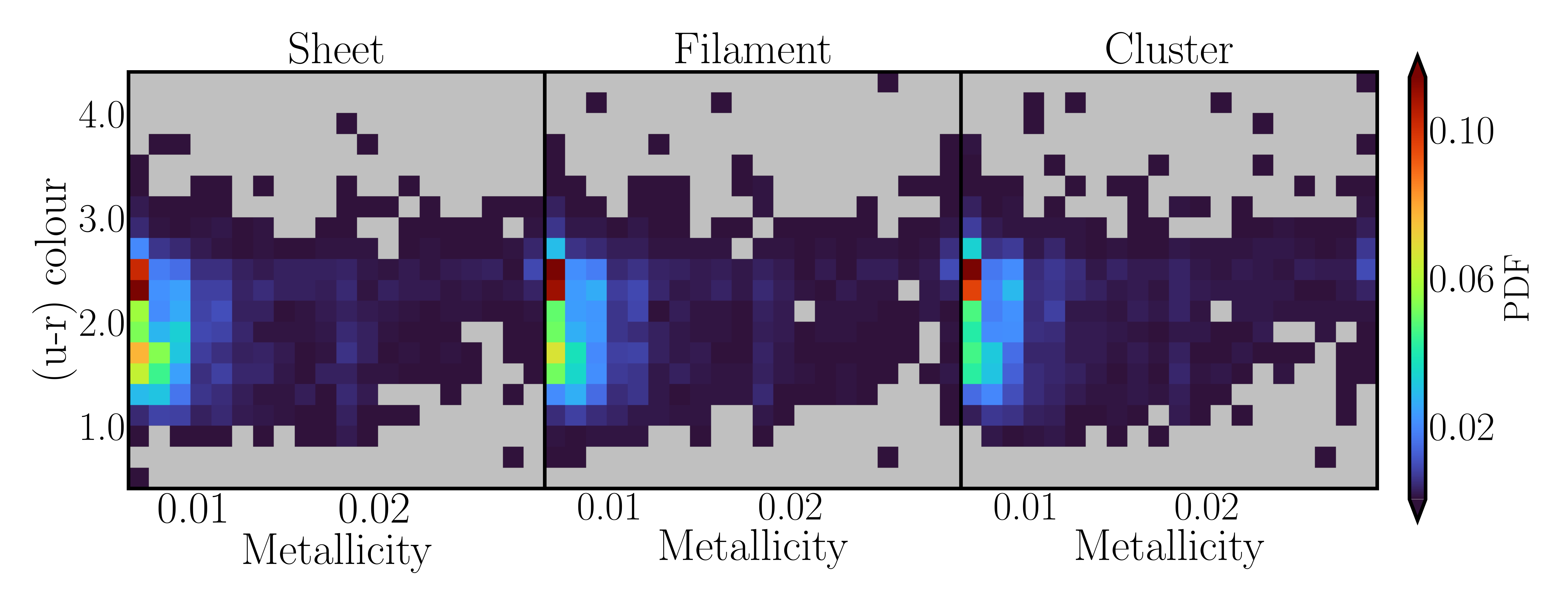}
\includegraphics[width=16cm]{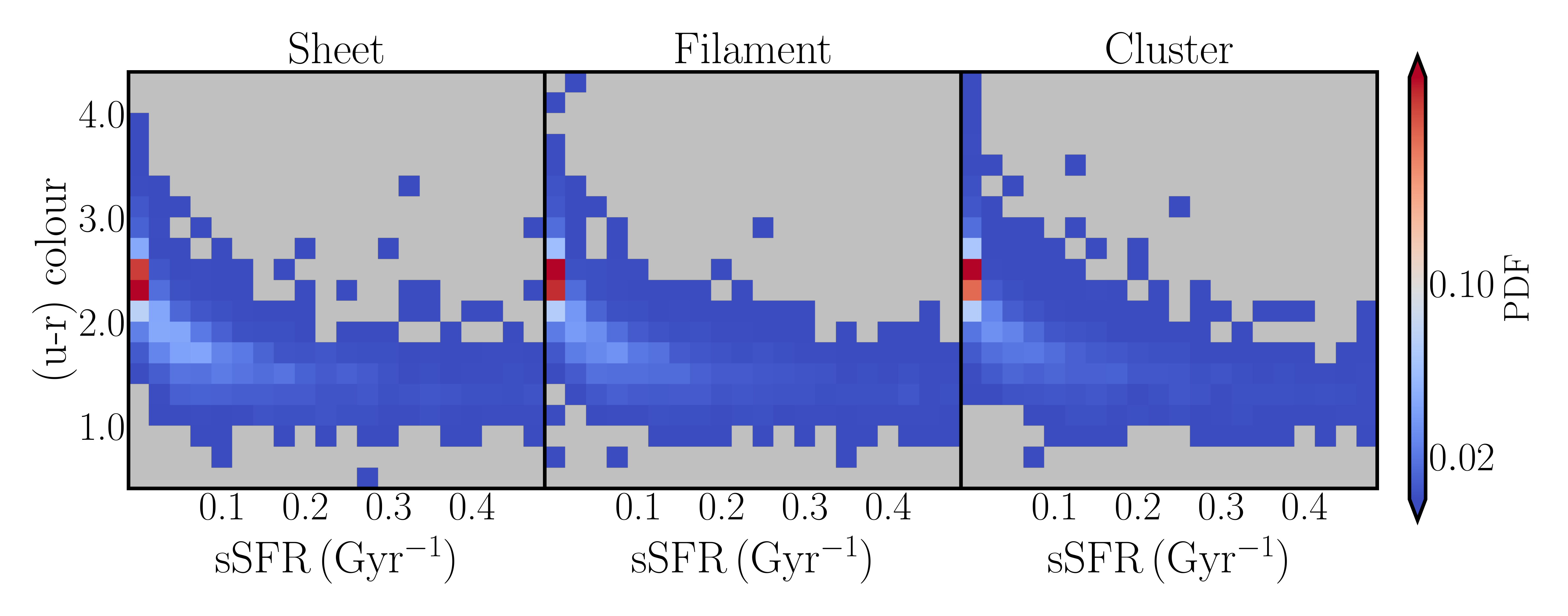}
\caption{The top, middle and bottom panels of this figure respectively
  show the joint probability distribution of $(u-r)$ colour with
  stellar mass, metallicity and sSFR in different geometric
  environments.}
\label{fig:6}
\end{figure}

\begin{figure}[h!]
\includegraphics[width=16cm]{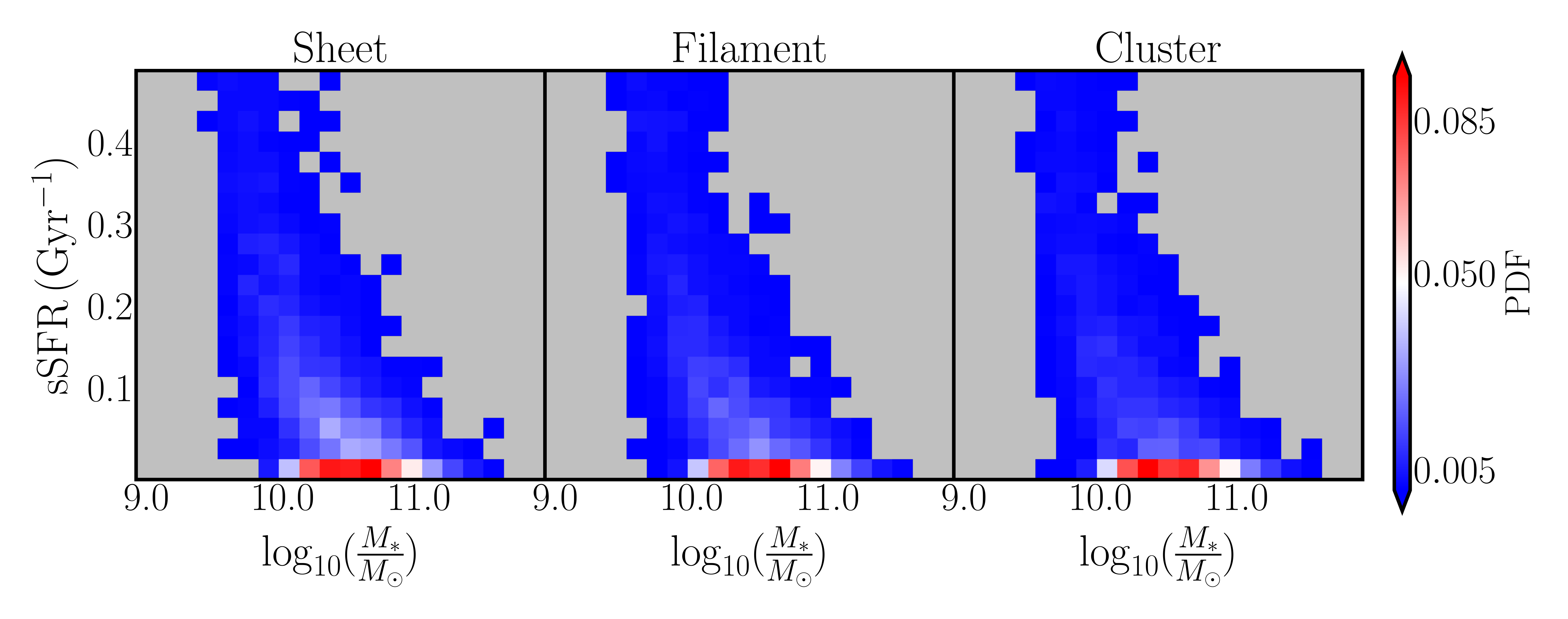}
\includegraphics[width=16cm]{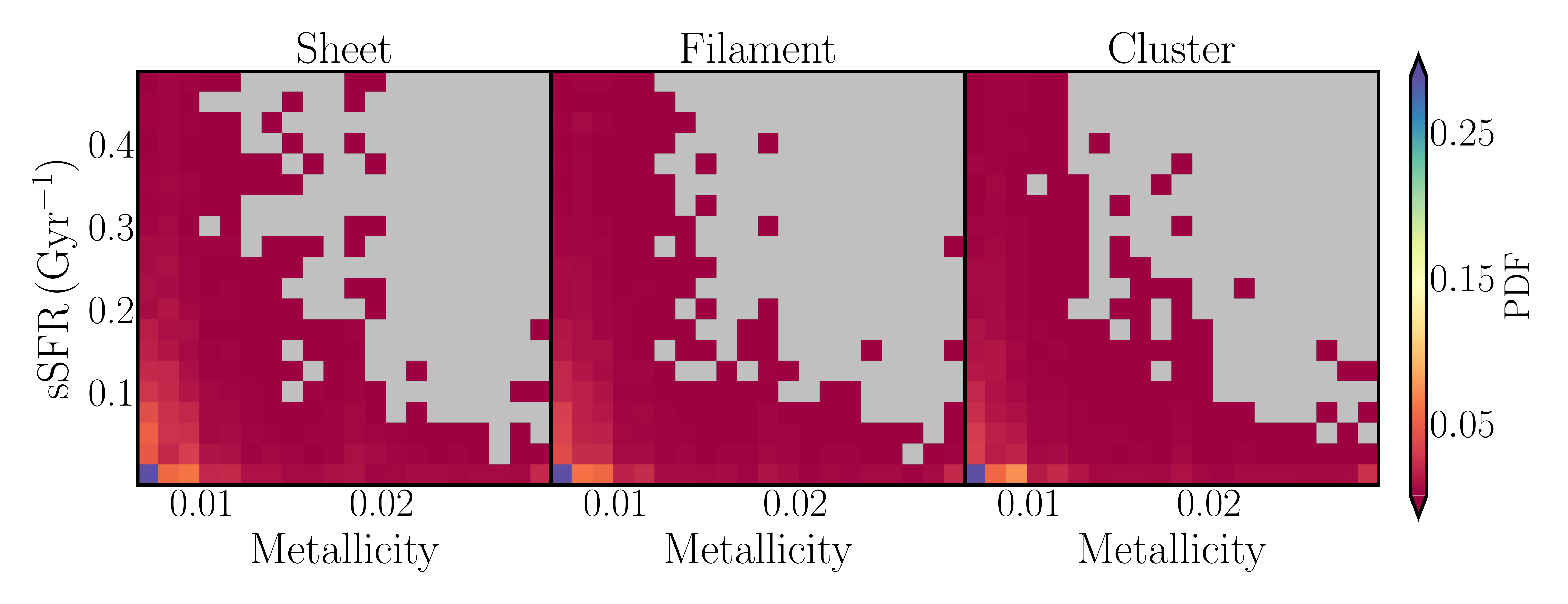}
\includegraphics[width=16cm]{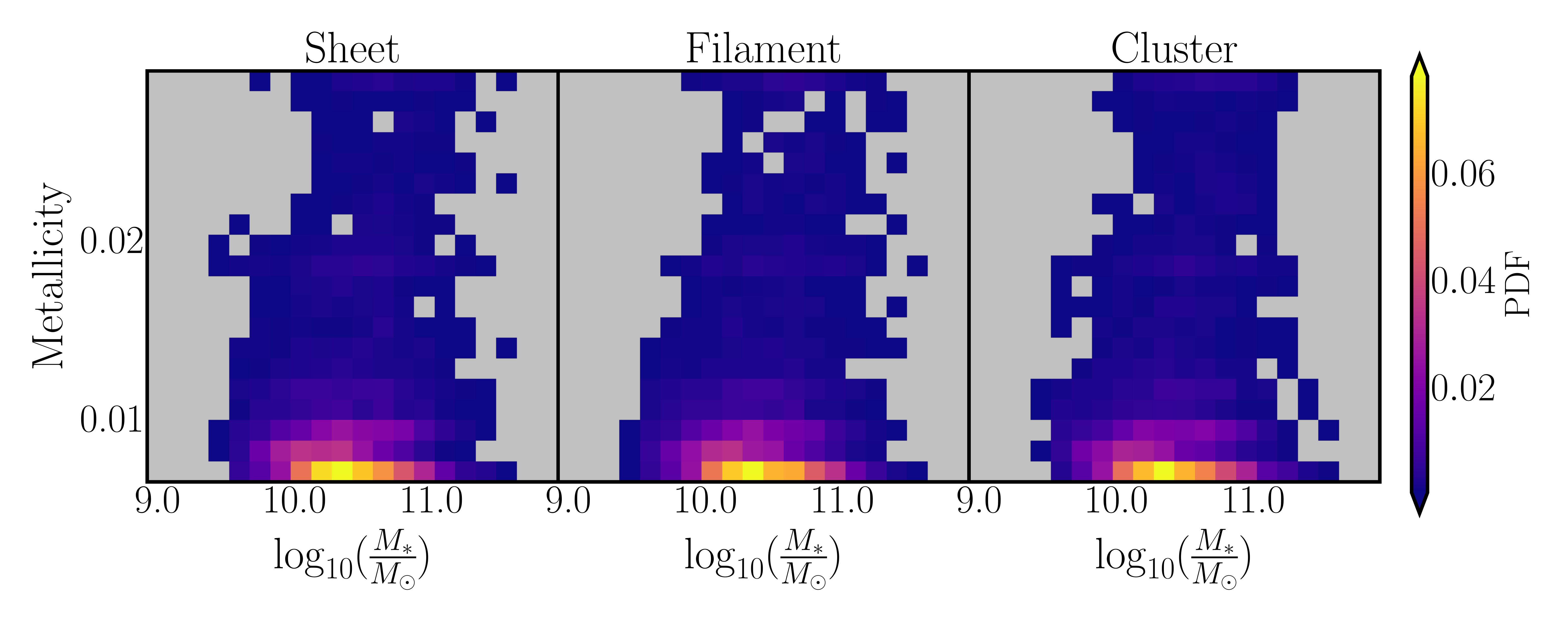}
\caption{Same as \autoref{fig:6} but for different pairs of galaxy
  properties.}
\label{fig:7}
\end{figure}

\subsection{Measuring the correlations with the Pearson correlation coefficient}
\label{sec:pearson}
In \autoref{sec:median}, we use \autoref{fig:3} to infer the relations
between different galaxy properties in a qualitative manner. One can
quantify the magnitudes and directions of these correlations by
measuring the PCC between each pair of galaxy properties in different
morphological environments.

We show the PCCs for different pairs of galaxy properties in each type
of cosmic web environments in the upper panels of \autoref{fig:4}. The
$1\sigma$ errors in each of these measurements are shown in the bottom
panels of the same figure. The $1\sigma$ errors are obtained from 50
jackknife samples drawn from the original datasets. In the top three
panels, we find that the sSFR shows anti-correlation with each of the
other three galaxy properties in all types of morphological
environments. Out of these three pairs, the sSFR-stellar mass and
sSFR-(u-r) colour are strongly anti-correlated whereas the
sSFR-metallicity is weakly anti-correlated. Metallicity shows
relatively stronger correlations with (u-r) colour and stellar
mass. The stellar mass is strongly correlated with the (u-r) colour in
all environments which partly explains the observed positive
correlations for the metallicity-stellar mass and metallicity-(u-r)
colour relations. The PCC for each pair of galaxy properties in each
environment are significantly larger compared to the $1\sigma$ errors
in the respective measurements. So the measured correlations and
anti-correlations between different galaxy properties are
statistically significant. The PCC corresponding to each pair of
galaxy properties differ by a small amount across the different
environments. Interestingly, the $1\sigma$ errors for the correlation
coefficients are also very small.


\begin{figure}[h!]
\centering \includegraphics[width=15cm]{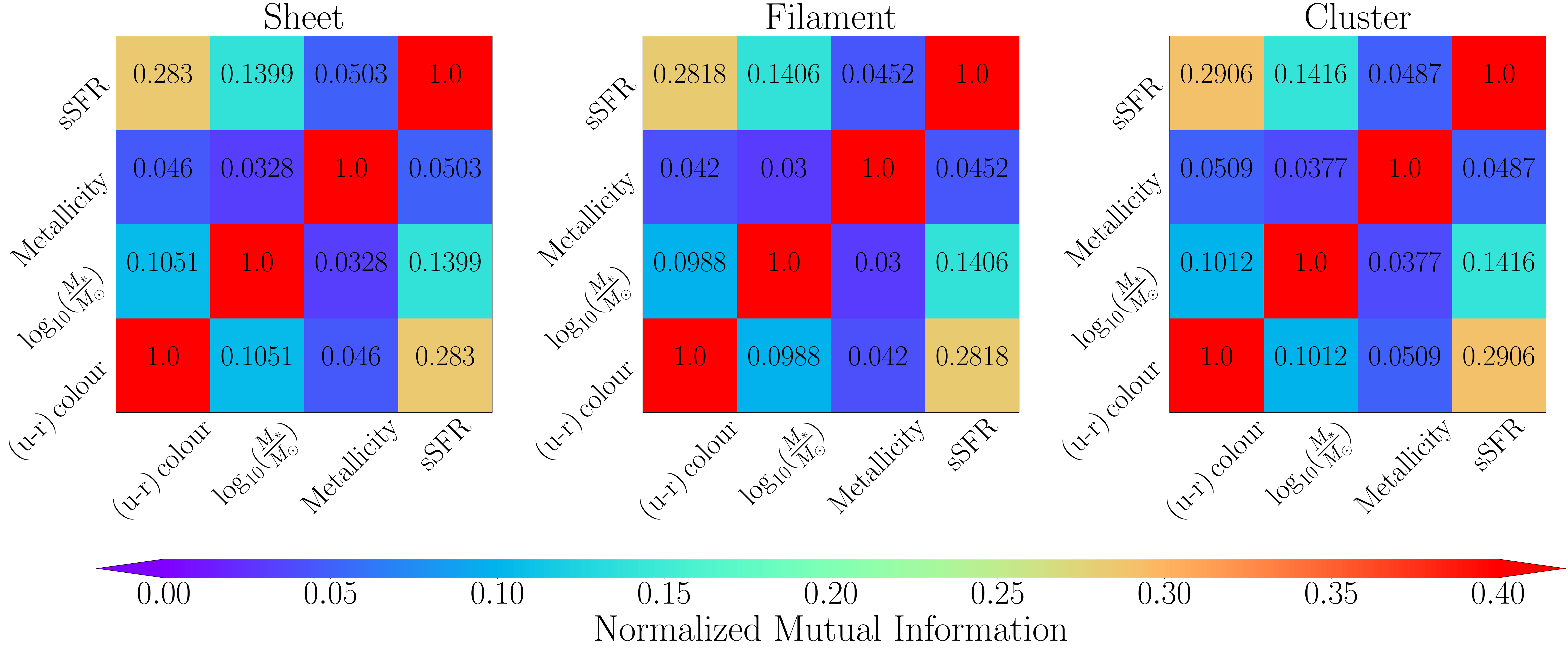}
\vspace{20px}

\includegraphics[width=15cm]{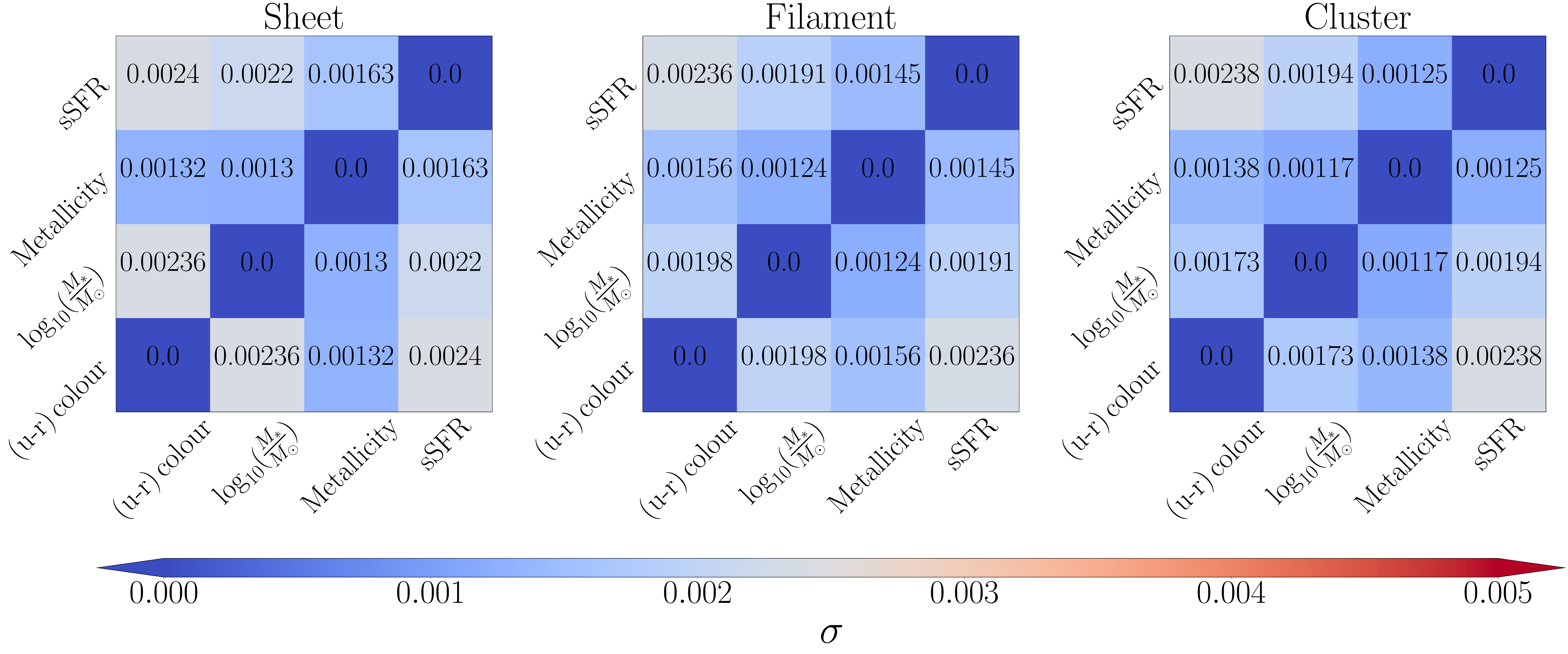}
\caption{Same as \autoref{fig:4} but for NMI. We choose 20 bins for this NMI analysis.}
\label{fig:8}
\end{figure}

We use a two-tailed t-test to determine if the observed PCC for
various pairs of galaxy properties are significantly different in
filaments, sheets and clusters. We compare the relations for separate
geometric environments. The t-score and p-value corresponding to each
relation are obtained in different pair of geometric environments and
listed in \autoref{tab:3}. We find a strong evidence against the null
hypothesis for most of the relations in different cosmic web
environments. However, the null hypothesis can not be rejected for the
colour-stellar mass, stellar mass-metallicity and sSFR-metallicity
relations in sheet and cluster. The same is true for the
colour-metallicity relation in sheet and filament. We note that the
majority of the relations are strongly sensitive to the geometric
environments of the cosmic web. It may be noted that these relations
are somewhat non-linear and non-monotonic (\autoref{fig:3}). The PCC
may not be ideally suited for measuring the correlations between these
galaxy properties. We address this limitation of PCC by employing NMI
for our analysis in the next subsection.

\begin{table}
\setlength\extrarowheight{-0.01pt}

\centering
\begin{tabular}{|c|c|c|c|c|c|c|}
\hline
{\rule{0pt}{4ex} Relations} & \multicolumn{2}{|c|}{ Sheet - Filament } & \multicolumn{2}{c|}{Filament - Cluster} & \multicolumn{2}{c|}{ Sheet-Cluster } \\ \cline{2-7} 
& $t$ score & $p$ value & $t$ score & $p$ value & $t$ score & $p$ value \\
\hline\hline
\rule{0pt}{3ex} colour-stellar mass & $13.83$ & $9.26 \times 10^{-25}$ & $-6.42$ & $4.79 \times 10^{-9}$ & $8.79$ & $5.09 \times 10^{-14}$ \\
\hline
\rule{0pt}{3ex} colour-metallicity & $13.46$ & $5.16 \times 10^{-24}$ & $-29.94$ & $4.13 \times 10^{-51}$ & $-18.24$ & $2.90 \times 10^{-33}$ \\
\hline
\rule{0pt}{3ex} colour-sSFR & $1.79$ & $7.68 \times 10^{-2}$ & $-19.69$ & $7.67 \times 10^{-36}$ & $-17.72$ & $2.55 \times 10^{-32}$ \\
\hline
\rule{0pt}{3ex} stellar mass-metallicity  & $11.26$ & $2.29 \times 10^{-19}$ & $-30.79$ & $3.48 \times 10^{-52}$ & $-18.40$ & $1.45 \times 10^{-33}$ \\
\hline
\rule{0pt}{3ex} stellar mass-sSFR & $-2.78$ & $6.54 \times 10^{-3}$ & $-1.93$ & $5.66 \times 10^{-2}$ & $-4.55$ & $1.53 \times 10^{-5}$ \\
\hline
\rule{0pt}{3ex} sSFR-metallicity & $17.08$ & $3.82 \times 10^{-31}$ & $-13.58$ & $2.90 \times 10^{-24}$ & $5.50$ & $3.08 \times 10^{-7}$\\
\hline
\end{tabular}
\caption{This table shows the $t$ scores and $p$ values calculated
  using Student's t-test after comparing the normalized mutual
  information between different pairs of galaxy properties in two
  separate cosmic web environments. We have used a two-tailed t-test for
  our analysis. The degrees of freedom in this test is $98$. 20 bins
  are used for this analysis.}
\label{tab:4}
\end{table}


\subsection{Quantifying the correlations using the normalized mutual information}

In \autoref{sec:pearson}, we use the PCC to measure the correlations
and anti-correlations between different galaxy properties in various
cosmic web environments. The PCC fails to capture any non-linear and
non-monotonic behaviour in these relationships. We see that the
relationship between the different galaxy properties are not entirely
linear (\autoref{sec:median}). So we also employ NMI to study the
correlations between these galaxy properties. This requires us to
calculate the probability distribution functions (PDF) for each galaxy
property as well as the 2D probability distribution functions
corresponding to each pair of galaxy properties.

We show the PDFs of (u-r) colour, sSFR and metallicity in the three
panels of \autoref{fig:5}. In the top left panel of \autoref{fig:5},
we find that the clusters host a larger fraction of the red galaxies
and a lower fraction of the blue galaxies. We see a similar trend in
filaments. Interestingly, the sheets host a comparable fraction of the
red and the blue galaxies. The top right panel of \autoref{fig:5}
shows the PDFs of sSFR in sheets, filaments and clusters. The PDFs
show that the galaxies hosted in the sheets and filaments are
relatively high star forming than those located in the
clusters. Similar differences can be noticed in the bottom panel of
\autoref{fig:5} which shows the PDFs of metallicity in different
morphological environments.

We show the joint probability distributions of different pairs of
galaxy properties in \autoref{fig:6} and \autoref{fig:7}. A closer
look at these plots suggests that there are similarities as well as
differences in these PDFs across the different cosmic environments.
We can see a distinct galaxy population with a very low sSFR and
metallicity that lies within
$10<\log_{10}(\frac{M_{\ast}}{M_{\odot}})<11$ and $2.25<(u-r)<2.75$ in
these plots. This population represents the quiescent galaxies that
are undergoing a transition due to either secular evolution or
environmental influences. A higher concentration of such galaxies in
clusters indicates that they provide a favourable environment for
quenching. It is clear that the transitional population is present
across all types of environments in the cosmic web.

We use these PDFs and the joint PDFs to calculate the NMI between
galaxy properties in different cosmic web environments. The NMI
between different galaxy properties in sheets, filaments and clusters
are shown in the top three panels of \autoref{fig:8}. We find a
non-zero NMI for each pair of galaxy properties in all the cosmic web
environments. The 1$\sigma$ errors in each of these measurements are
also shown in the bottom three panels of the same figure. The errors
are estimated from 50 jackknife samples drawn from the original
datasets.

In the top three panels of \autoref{fig:8}, we find that the sSFR
shows a relatively stronger correlation with colour and stellar mass
and a weaker correlation with metallicity. The correlation of
metallicity with colour and stellar mass are also relatively weaker in
all three environments. The correlation between colour and stellar
mass is reasonably stronger in all environments. It may be noted that
NMI does not provide the direction for any of these
relationships. However it is a better measure of correlations in
presence of non-linearity and non-Gaussianity. We use a two-tailed
t-test in order to asses the statistical significance of the
differences in the NMI in different geometric environments. We list
the t-scores and p-values corresponding to each relation for each pair
of geometric environments in \autoref{tab:4}. The p-values suggest
that the null hypothesis can be rejected for most of the relations in
different pairs of geometric environments. We note that there are
weaker evidences against the null hypothesis for the stellar mass-sSFR
relation in filament and cluster. Also, the null hypothesis can not be
rejected for the colour-sSFR relation in sheet and filament.

We use the same number of galaxies (4363) and the same number of bins
(20) across all the geometric environments in our NMI analysis. The
specific choice of bins may have an influence on our results.  We
consider the effects of binning by repeating the NMI analysis for two
other choices of the number of bins (15 and 25). The corresponding
results are shown in \autoref{fig:nmi_15} and \autoref{fig:nmi_25}
respectively. The values of the t-scores and p-values for the two
cases are listed in \autoref{tab:nmi_ttest_15_bins} and
\autoref{tab:nmi_ttest_25_bins}.  We find that the null hypothesis can
be rejected for most of the relations irrespective of our choice of
the number of bins. Our conclusions remain unchanged, and hence are
robust against the choice of the number of bins.

The analysis with PCC and NMI agree on discarding the null hypothesis
for most of the relations. There are a few differences too. The null
hypothesis for the colour-sSFR relation in sheet-filament and the
stellar mass-sSFR relation in filament-cluster can be rejected if PCC
is used instead of NMI. On the other hand, the NMI analysis allows us
to reject the null hypothesis for which PCC provides weak evidences
(e.g. colour-stellar mass, stellar mass-metllicity relations in
Sheet-Cluster). Given the non-linear nature of the relationships, the
NMI should be preferred over the PCC in breaking any such
ambiguities. Finally, the results of our analysis clearly indicate
that the correlations between different galaxy properties are strongly
sensitive to the geometric environments of the cosmic web.

\section{Conclusions}

We analyze the correlations between $(u-r)$ colour, stellar mass, sSFR
and metallicity in different geometric environments of the cosmic
web. The geometric environments of the galaxies are identified using
the eigenvalues of the 3D tidal tensor. We use Pearson correlation
coefficient (PCC) and the normalized mutual information to study the
correlations. The PCC is a well known measure of correlation that
assumes Gaussianity in the distributions of the variables and
linearity in their relationships. However, this may not hold true for
each galaxy property and every relationship. The NMI does not make any
assumption regarding the distribution of the variables and the nature
of relationship between them. It is positive by definition and hence
can not capture the direction in a relation unlike the PCC. Its
primary advantage is that it can efficiently capture the correlations
even when the relationships are non-linear and non-monotonic. We use
the NMI as a complimentary measure of correlation between the galaxy
properties. We compare the observed PCC and NMI between the galaxy
properties across the different geometric environments. Using a
two-tailed t-test, we investigate if the mean PCC and NMI in sheet,
filament and cluster are significantly different. Both the PCC and the
NMI allow us to reject the null hypothesis for most of the relations
in different environments. The null hypothesis could not be rejected
in only two instances out of eighteen while using NMI and four cases
out of eighteen while using PCC. We believe that NMI is a better
choice for our study due to the non-linear nature of the
relationships. Our results suggest that most of the scaling relations
are strongly sensitive to the cosmic web environment.

The relationship between the galaxy properties and the environment is
a complex issue. Galaxies follow different evolutionary pathways. Both
the local environment and the geometry of the cosmic web may have
significant influences on the evolution of galaxy properties. Any such
dependence may also modulate the scaling relations in different
environments. Studies with the hydrodynamical simulations suggest that
the cosmic web dependence of the galaxy properties is primarily driven
by the variation of halo mass functions with environment
\citep{metuki15, xu20}. Some earlier studies with the SDSS data find
that the galaxy properties have a weak dependence on the cosmic web
environments \citep{paranjape18, alam19, kuutma20}. On the other hand,
several other works demonstrate the roles of the cosmic web in
deciding the galaxy properties \citep{scudder, lupa, pandey2, pandey3,
  darvish, filho}.  The galaxy properties are also known to be
influenced by the tidal environments \citep{erdogdu, tempel1}. The
spin and shape of the dark matter halos are known to have alignment
with their host environments \citep{wang18, veena18}. Some recent
studies with simulations and observations show that the cosmic
filaments have spin on unprecedented scales \citep{xia21,
  wang21}. Using the SDSS data, \cite{kraljic20} show that galaxy
properties exhibit clear dependence on the connectivity of the cosmic
web at fixed stellar mass. Such dependence on the connectivity of the
cosmic web suggests that it has crucial roles in providing the main
source of fuel for galaxy growth. An analysis of the IllustrisTNG
simulation \citep{vogelsberger} by \cite{hasan23} find significant
dependence of galaxy properties on the cosmic web environment. They
show that the cosmic web structures efficiently channel cold gas into
most galaxies leading to such dependence. Our results are consistent
with these findings. These together suggest that the large-scale
geometric environments have important roles in the galaxy evolution.

Nevertheless, the interpretation of our results may be affected by a
number of issues related to the density and structural variety of the
structures. The various cosmic web environments represent the regions
of different overdensity. The clusters are generally denser than the
filaments and the filaments are relatively denser than the sheets. The
local density and the geometric environments are expected to have
combined influences on the galaxy properties and their correlations.
It is difficult to separate these influences in our study. We tried to
carry out an analysis by restricting the local density within a fixed
range for each environment. However, this provides us with a very
small number of galaxies at each environment. There are other subtle
issues that are not addressed in the present work. For example, the
filaments around the massive clusters may have different density and
morphology than filaments around the low-mass clusters. The filaments
may have different visual morphology such as long, short, straight,
curved, warped etc \citep{pimb04}. The environmental influences on
galaxies in massive clusters and low mass clusters are also expected
to be different. It would be interesting to study these relationships
in structures of different mass and visual morphology. These may be
considered as some caveats in our analysis. We plan to study these
issues in greater detail in a future study using the Illustris
simulations \citep{vogelsberger}. Further, the sensitivity of the
correlations to the geometric environment may depend on the details of
the galaxy formation models as well as the background cosmological
models. It would be interesting to investigate these issues using
simulations.

Finally, we conclude from our study that the correlations between
different galaxy properties are susceptible to the geometric
environments of the cosmic web.

\section*{ACKNOWLEDGEMENT}
The authors thank an anonymous reviewer for the insightful comments
and suggestions that helped to improve the draft. AN acknowledges the
financial support from the Department of Science and Technology (DST),
Government of India through an INSPIRE fellowship. BP would like to
acknowledge financial support from the SERB, DST, Government of India
through the project CRG/2019/001110. BP would also like to acknowledge
IUCAA, Pune, for providing support through the associateship
programme.

Funding for the SDSS and SDSS-II has been provided by the Alfred
P. Sloan Foundation, the Participating Institutions, the National
Science Foundation, the U.S. Department of Energy, the National
Aeronautics and Space Administration, the Japanese Monbukagakusho, the
Max Planck Society, and the Higher Education Funding Council for
England. The SDSS website is http://www.sdss.org/.

The SDSS is managed by the Astrophysical Research Consortium for the
Participating Institutions. The Participating Institutions are the
American Museum of Natural History, Astrophysical Institute Potsdam,
University of Basel, University of Cambridge, Case Western Reserve
University, University of Chicago, Drexel University, Fermilab, the
Institute for Advanced Study, the Japan Participation Group, Johns
Hopkins University, the Joint Institute for Nuclear Astrophysics, the
Kavli Institute for Particle Astrophysics and Cosmology, the Korean
Scientist Group, the Chinese Academy of Sciences (LAMOST), Los Alamos
National Laboratory, the Max-Planck-Institute for Astronomy (MPIA),
the Max-Planck-Institute for Astrophysics (MPA), New Mexico State
University, Ohio State University, University of Pittsburgh,
University of Portsmouth, Princeton University, the United States
Naval Observatory, and the University of Washington.

\appendix 
\section{Appendix}
\label{sec:appen}
\begin{figure}[h!]
\centering \includegraphics[width=15cm]{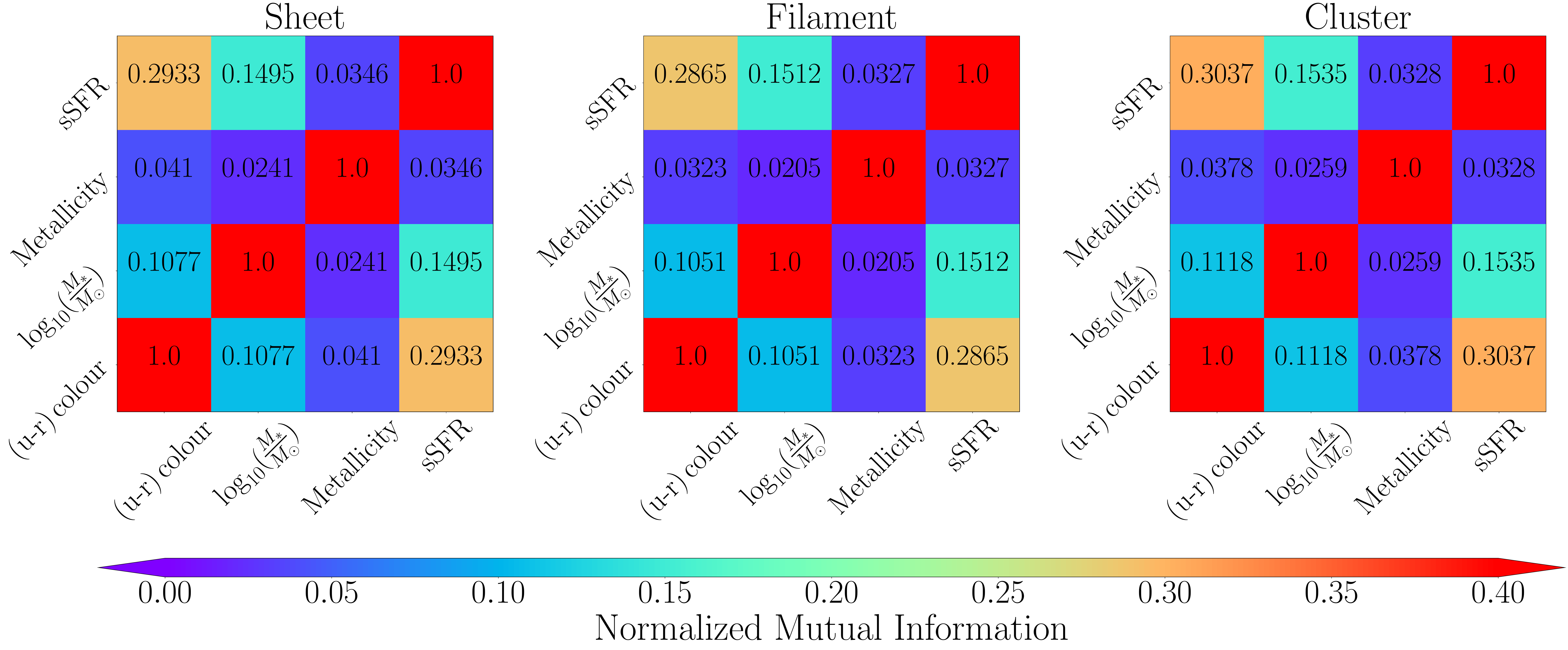}
\vspace{20px}

\includegraphics[width=15cm]{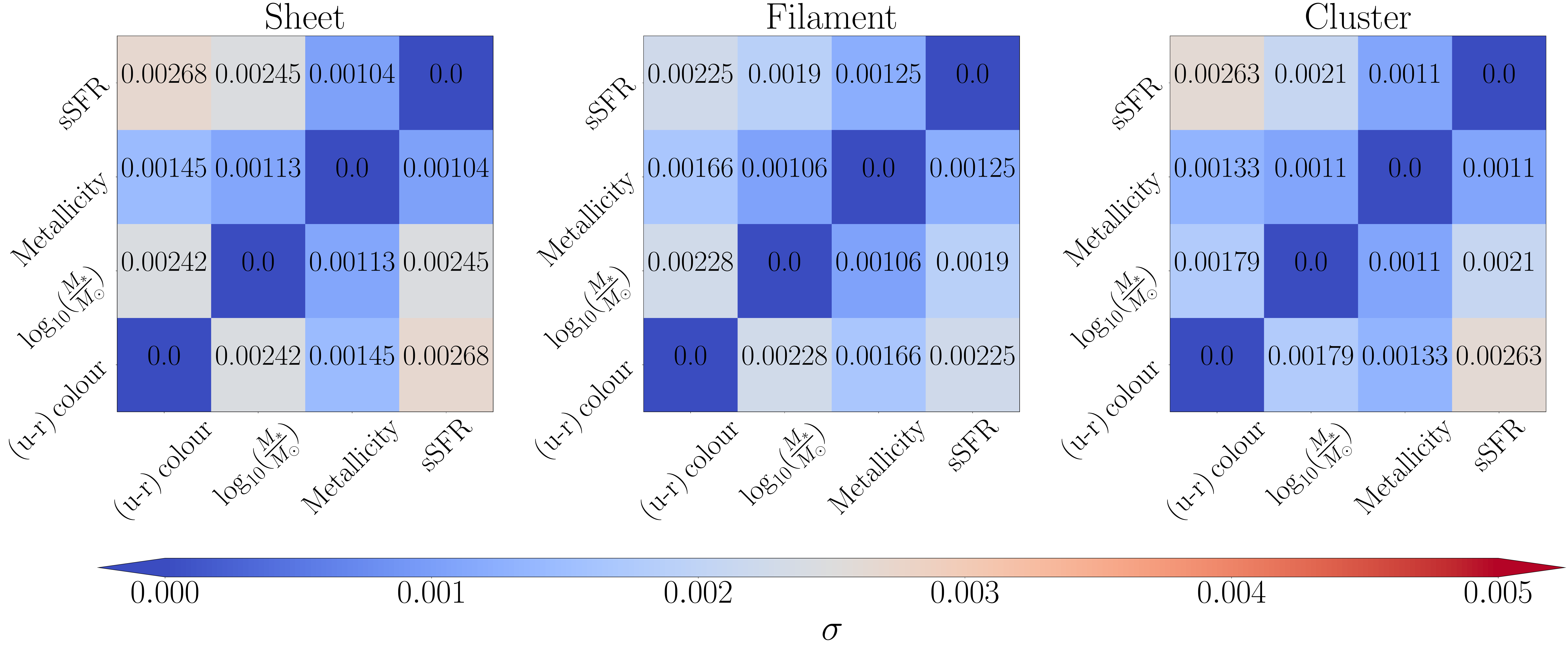}
\caption{Same as \autoref{fig:8} but for 15 bins.}
\label{fig:nmi_15}
\end{figure}

\begin{figure}[h!]
\centering \includegraphics[width=15cm]{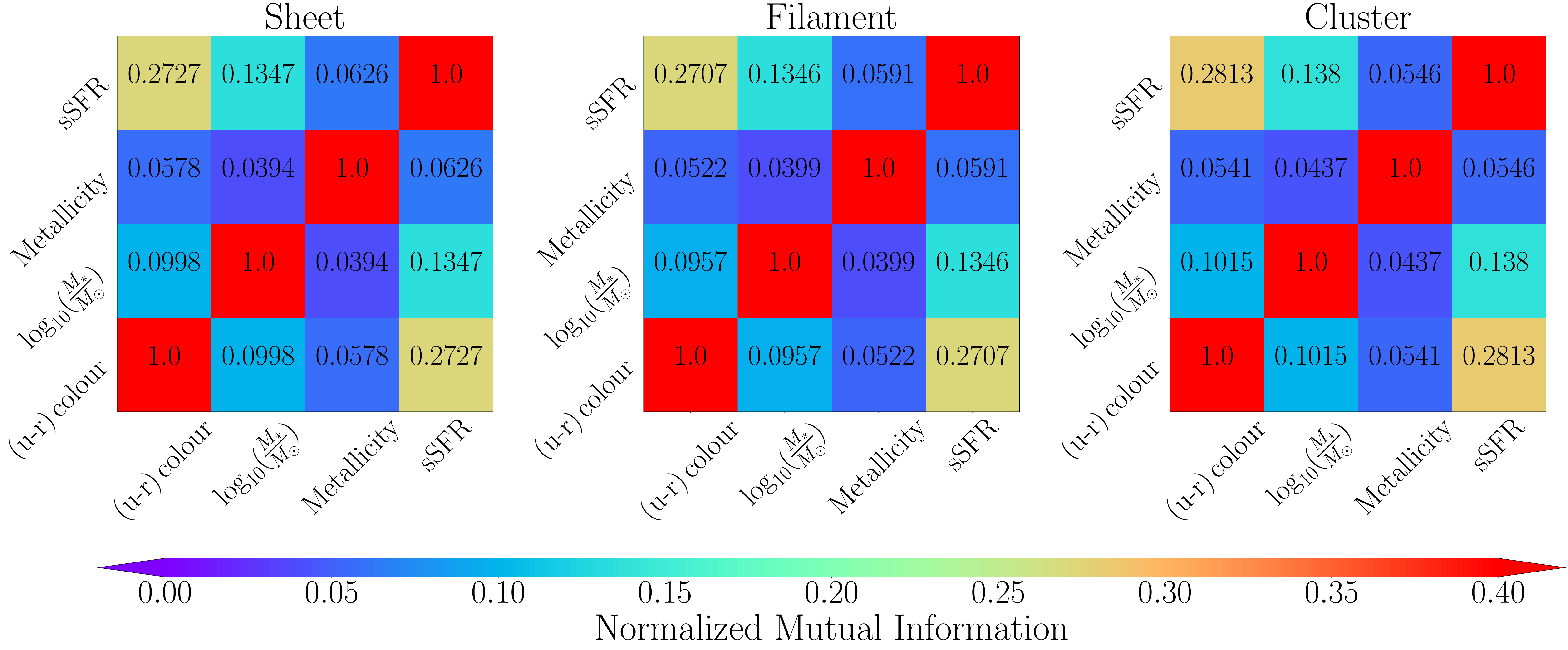}
\vspace{20px}

\includegraphics[width=15cm]{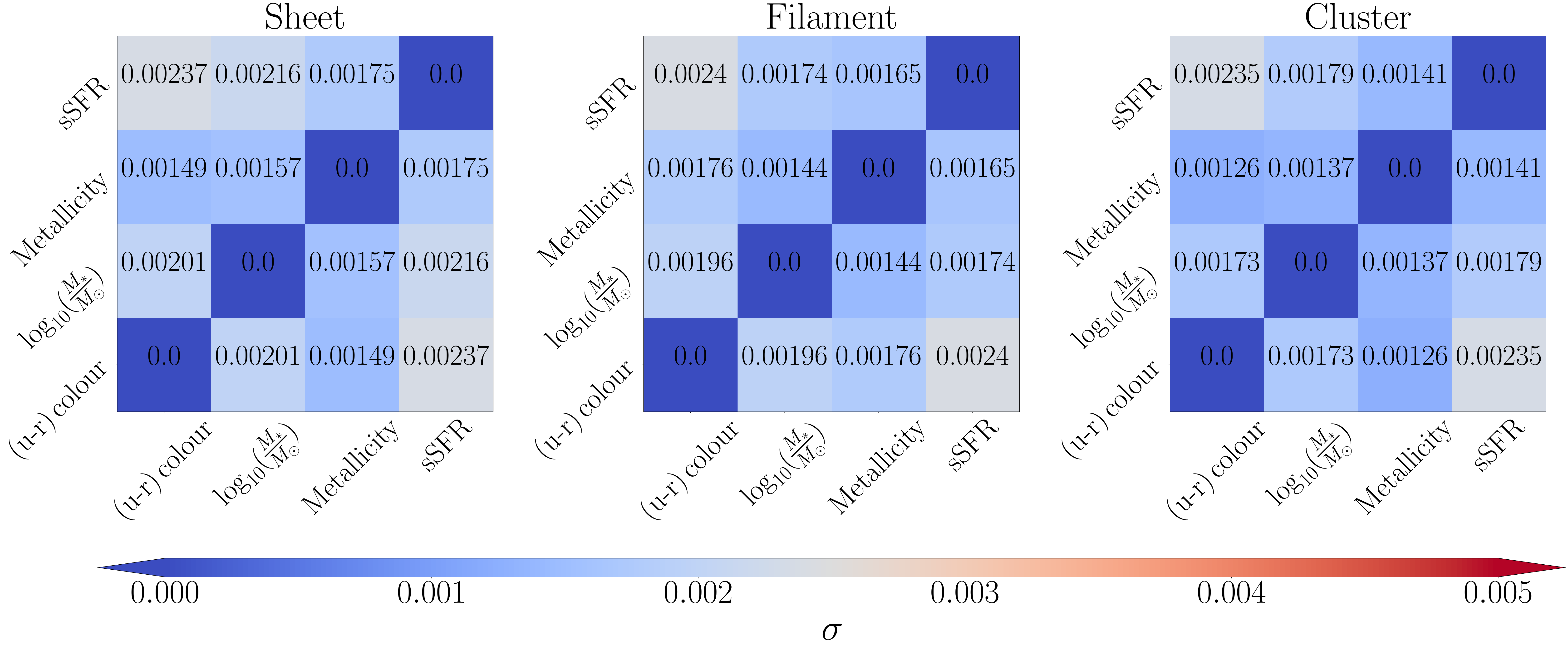}
\caption{Same as \autoref{fig:8} but for 25 bins.}
\label{fig:nmi_25}
\end{figure}

We repeat our NMI analysis for 15 bins and 25 bins. The respective
results are shown in \autoref{fig:nmi_15} and \autoref{fig:nmi_25}
respectively. We list the t-scores and p-values for the two cases in
\autoref{tab:nmi_ttest_15_bins} and \autoref{tab:nmi_ttest_25_bins}.

\begin{table}
\centering
\begin{tabular}{|c|c|c|c|c|c|c|}
\hline
{\rule{0pt}{4ex} Relations}& \multicolumn{2}{|c|}{ Sheet - Filament } & \multicolumn{2}{c|}{Filament - Cluster} & \multicolumn{2}{c|}{ Sheet-Cluster } \\ \cline{2-7} 
& $t$ score & $p$ value & $t$ score & $p$ value & $t$ score & $p$ value \\
\hline\hline
\rule{0pt}{3ex} colour-stellar mass & $4.78$ & $6.08 \times 10^{-6}$ & $-15.38$ & $7.02 \times 10^{-28}$ & $-9.51$ & $1.41 \times 10^{-15}$ \\
\hline
\rule{0pt}{3ex} colour-metallicity & $28.06$ & $1.24 \times 10^{-48}$ & $-19.89$ & $3.54 \times 10^{-36}$ & $9.94$ & $1.62 \times 10^{-16}$ \\
\hline
\rule{0pt}{3ex} colour-sSFR & $14.99$ & $4.17 \times 10^{-27}$ & $-37.07$ & $1.82 \times 10^{-59}$ & $-20.22$ & $9.55\times10^{-37}$ \\
\hline
\rule{0pt}{3ex} stellar mass-metallicity  & $16.14$ & $2.33\times10^{-29}$ & $-24.97$ & $2.88 \times 10^{-44}$ & $-8.27$ & $6.59 \times 10^{-13}$ \\
\hline
\rule{0pt}{3ex} stellar mass-sSFR & $-4.85$ & $4.61 \times 10^{-6}$ & $-4.57$ & $1.44 \times 10^{-5}$ & $-8.66$ & $9.59 \times 10^{-14}$ \\
\hline
\rule{0pt}{3ex} sSFR-metallicity & $8.50$ & $2.12 \times 10^{-13}$ & $-0.75$ & $4.58 \times 10^{-1}$ & $8.32$ & $5.20 \times 10^{-13}$\\
\hline
\end{tabular}
\caption{Same as \autoref{tab:4} but for 15 bins.}
\label{tab:nmi_ttest_15_bins}
\end{table}

\newpage
\begin{table}
\centering
\begin{tabular}{|c|c|c|c|c|c|c|}
\hline
{\rule{0pt}{4ex} Relations}& \multicolumn{2}{|c|}{ Sheet - Filament } & \multicolumn{2}{c|}{Filament - Cluster} & \multicolumn{2}{c|}{ Sheet-Cluster } \\ \cline{2-7} 
& $t$ score & $p$ value & $t$ score & $p$ value & $t$ score & $p$ value \\
\hline\hline
\rule{0pt}{3ex} colour-stellar mass & $10.01$ & $1.13 \times 10^{-16}$ & $-15.58$ & $2.87 \times 10^{-28}$ & $-4.77$ & $6.49 \times 10^{-6}$ \\
\hline
\rule{0pt}{3ex} colour-metallicity & $16.66$ & $2.41 \times 10^{-30}$ & $-6.62$ & $1.94 \times 10^{-9}$ & $12.33$ & $1.22 \times 10^{-21}$ \\
\hline
\rule{0pt}{3ex} colour-sSFR & $3.59$ & $5.19 \times 10^{-4}$ & $-23.08$ & $2.06 \times 10^{-41}$ & $-19.59$ & $1.17 \times 10^{-35}$ \\
\hline
\rule{0pt}{3ex} stellar mass-metallicity  & $-1.57$ & $1.21 \times 10^{-1}$ & $-13.51$ & $4.21 \times 10^{-24}$ & $-14.51$ & $3.74 \times 10^{-26}$ \\
\hline
\rule{0pt}{3ex} stellar mass-sSFR & $-0.85$ & $3.97 \times 10^{-1}$ & $-8.32$ & $5.37 \times 10^{-13}$ & $-8.24$ & $7.97 \times 10^{-13}$ \\
\hline
\rule{0pt}{3ex} sSFR-metallicity & $10.06$ & $9.09 \times 10^{-17}$ & $12.25$ & $1.75 \times 10^{-21}$ & $22.59$ & $1.25 \times 10^{-40}$\\
\hline
\end{tabular}
\caption{Same as \autoref{tab:4} but for 25 bins.}
\label{tab:nmi_ttest_25_bins}
\end{table}
\end{document}